\documentclass{aa}  
\bibpunct{(}{)}{;}{a}{}{,} 
\usepackage{graphicx}
\usepackage{siunitx}
\DeclareSIUnit \parsec {pc}

\newcommand{\um}[1]{\SI{#1}{\micro\meter}}

\usepackage{txfonts}
\usepackage[svgnames]{xcolor}
\definecolor{bluegrey}{RGB}{42, 54, 112}
\usepackage[colorlinks=true, citecolor=bluegrey, linkcolor=bluegrey]{hyperref}
\begin{document}

\title{Predicting the global far-infrared SED of galaxies \\
via machine learning techniques}

\author{W. Dobbels
	\inst{1}
	\and M. Baes\inst{1}
	\and S. Viaene\inst{1, 2}
	\and S. Bianchi\inst{3}
	\and J.~I.~Davies\inst{4}
	\and V.~Casasola\inst{3,5}
	\and C.~J.~R.~Clark\inst{6}
	\and J.~Fritz \inst{7}
	\and M.~Galametz\inst{8}
	\and F. Galliano\inst{8}	
	\and A.~Mosenkov\inst{1,9, 10}
	\and A.~Nersesian\inst{1,11, 12}
	\and A.~Tr\v{c}ka\inst{1}
}

\institute{Sterrenkundig Observatorium, Universiteit Gent, Krijgslaan 281, B-9000 Gent, Belgium \\
	\email{wouter.dobbels@ugent.be}
	\and
	Centre for Astrophysics Research, University of Hertfordshire, College Lane, Hatfield AL10 9AB, UK
	\and
	INAF -- Osservatorio Astrofisico di Arcetri, Largo E. Fermi 5, 50125 Florence, Italy
	\and
	School of Physics \& Astronomy, Cardiff University, Queen’s Buildings, The Parade, Cardiff, CF24 3AA, UK
	\and
	INAF -- Istituto di Radioastronomia, Via P. Gobetti 101, 4019 Bologna, Italy
	\and
	Space Telescope Science Institute, 3700 San Martin Drive, Baltimore, Maryland 21218, USA
	\and
	Instituto de Radioastronomía y Astrofísica, UNAM, Campus Morelia, A.P. 3-72, C.P. 58089, Mexico
	\and
	AIM, CEA, CNRS, Université Paris-Saclay, Université Paris Diderot, Sorbonne Paris Cité, 91191 Gif-sur-Yvette, France
	\and
	Central Astronomical Observatory of RAS, Pulkovskoye Chaussee 65/1, 196140 St. Petersburg, Russia
	\and
	St. Petersburg State University, Universitetskij Pr. 28, 198504 St. Petersburg, Stary Peterhof, Russia
	\and
	National Observatory of Athens, Institute for Astronomy, Astrophysics, Space Applications and Remote Sensing, Ioannou Metaxa and Vasileos Pavlou, 15236 Athens, Greece
	\and
	Department of Astrophysics, Astronomy \& Mechanics, Faculty of Physics, University of Athens, Panepistimiopolis, 15784 Zografos, Athens, Greece
}

\date{Accepted by A\&A, October 4, 2019}

\abstract
{Dust plays an important role in shaping a galaxy's spectral energy distribution (SED). It absorbs ultraviolet (UV) to near-infrared (NIR) radiation and re-emits this energy in the far-infrared (FIR). The FIR is essential to understand dust in galaxies. However, deep FIR observations require a space mission, none of which are still active today.}
{We aim to infer the FIR emission across six \textit{Herschel} bands, along with dust luminosity, mass, and effective temperature, based on the available UV to mid-infrared (MIR) observations. We also want to estimate the uncertainties of these predictions, compare our method to energy balance SED fitting, and determine possible limitations of the model.}
{We propose a machine learning framework to predict the FIR fluxes from 14 UV--MIR broadband fluxes. We used a low redshift sample by combining DustPedia and H-ATLAS, and extracted Bayesian flux posteriors through SED fitting. We trained shallow neural networks to predict the far-infrared fluxes, uncertainties, and dust properties. We evaluated them on a test set using a root mean square error (RMSE) in log-space.}
{Our results (RMSE = 0.19~dex) significantly outperform UV--MIR energy balance SED fitting (RMSE = 0.38~dex), and are inherently unbiased. We can identify when the predictions are off, for example when the input has large uncertainties on WISE \um{22}, or when the input does not resemble the training set. }
{The galaxies for which we have UV--FIR observations can be used as a blueprint for galaxies that lack FIR data. This results in a `virtual FIR telescope', which can be applied to large optical-MIR galaxy samples. This helps bridge the gap until the next FIR mission.}

\keywords{Galaxies: photometry -- Galaxies: ISM -- Infrared: galaxies
}

\maketitle
%

\section{Introduction}

Far-infrared (FIR) radiation is a key ingredient to study dust in galaxies. While dust manifests itself in ultraviolet (UV) and optical radiation through attenuation, it only emits at longer wavelengths \citep{dust-review}. This dust emission is important to estimate most dust properties, such as dust mass and temperature \citep{ciesla2014, cortese2012, auld2013, skibba2011}. In energy balance spectral energy distribution (SED) fitting, FIR helps to constrain both dust and stellar properties \citep{sedfit-review-conroy, dustpedia-cigale, smith2012-hatlas, malek2018-help}. These studies find that about one third of starlight in spiral galaxies is reprocessed by dust \citep{viaene2016, fabs}. A UV--submm SED is also required to study the star-dust interaction with radiative transfer models \citep{delooze2014-rt, viaene2017-rt, nersesian-rt}.

Unfortunately, observing in the FIR (here defined from \um{70} to \um{500}) poses some problems. Ground observations are limited to the longest wavelengths \citep[e.g. SCUBA-2;][]{scuba2}. Airborne telescopes are possible \citep[e.g. SOFIA;][]{sofia}, but in order to reach the highest sensitivities, a space mission is preferred. The instruments need to be cooled to observe the longer wavelengths, which limits the mission's lifetime. At the diffraction limit, the resolution is about a factor of 200 worse than in the optical, which leads to confused sources \citep{herschel-confusion, herschel-confusion-2}. The last FIR space mission, the \textit{Herschel} Space Observatory \citep{herschel}, ended in 2013. Possible successors, such as SPICA \citep{spica} and Origins \citep{origins}, have been proposed but not confirmed, and will launch no sooner than 2032.

The UV--near-infrared (NIR) radiation is related to the FIR radiation through energy balance: in thermal equilibrium, the energy absorbed by dust at shorter wavelengths is re-emitted at longer wavelengths. This total reprocessed energy is directly inferable from the FIR. In the absence of FIR detections, this total energy can only be constrained by assuming the shape of the unattenuated SED and an attenuation law. Here, SED modelling can provide an answer \citep[see][]{sedfit-review-walcher, sedfit-review-conroy}. A library of SED models is built by assuming a star-formation history (SFH), simple stellar population (SSP), and attenuation law. For each model, a $\chi^2$ is calculated, which determines how well the model fits the observed fluxes, under the assumption of uncorrelated errors. Through a Bayesian analysis, intrinsic parameters can be estimated \citep{cigale, magphys, pcigale, leja2017, beagle, bagpipes}. An observed UV--NIR SED can lead to an estimate of the total absorbed (and hence emitted) energy. There can, however, be some degeneracy between reddening from dust, and the reddening from an older and/or more metal-rich population. Moreover, energy balance does not hold for individual viewing angles: the FIR is isotropic but the UV--NIR is not, leading to more attenuation for an edge-on view.

Besides estimating the total emitted FIR energy, we might want to go a step further. The question is begged as to whether we can estimate a more detailed shape of the FIR spectrum from the UV--MIR radiation. In our SED models, we can include a dust emission model, and estimate FIR broadband fluxes. While the total emitted energy is constrained by the energy balance, the shape is not, and emission properties are varied independently from the absorption properties. This results in a single optical fit that is able to produce a wide range of FIR SEDs. Depending on the data set and the model's assumptions, these predictions achieve uncertainties between 0.2~dex and 1~dex \citep[][see also Sect. \ref{sec-sedfit}]{safar-fir-predict, chang-fir-predict, leja2017}.

Another approach to estimate the FIR is to make a full 3D model of the galaxy, and fit this using inverse radiative transfer to the UV--MIR radiation \citep{en-bal-baes, en-bal-delooze, en-bal-degeyter, en-bal-mosenkova, en-bal-mosenkovb}. This then produces a FIR spectrum, but this is usually about a factor of three below the measured values. It is believed that the cause of this discrepancy is due to the inability to resolve smaller dust clumps, such as in star-forming regions \citep{en-bal-saftly}.

In this work, we avoid the difficulties in explicitly modelling the star-dust interaction through the use of machine learning (ML). The goal is to build a predictor that takes the global UV--MIR SED as input, and produces an estimate of the global FIR fluxes in the six \textit{Herschel} broadbands. Of course, unlike the modelling, this can not teach us about all of the intricacies of the star-dust interaction. However, it does have some benefits. Given enough data, a machine learning approach can outperform models (if necessary by learning the differential). Machine learning also works on any kind of data, and there is no need for a different model for different classes of galaxies (star-forming and quiescent, low and high redshift, ...). Since our main focus is on FIR fluxes, and since we do not use physical models, we do not suffer from possible biases in these models. The machine learning model can make use of underlying correlations between stellar and dust properties. These correlations can be the result of galaxy evolution; for example, dust in early type galaxies tends to be hotter than in late type galaxies \citep{dust-etg, dustpedia-cigale}.

The goals of this paper are as follows. We predict the FIR SED, based on a UV--NIR SED. If accurate enough, the ML predictions can be used as virtual observations for galaxies that lack FIR data. In order to realise this, we also estimate the uncertainties on the individual predictions. We go one step beyond the fluxes, and also estimate the dust luminosity, mass, and temperature. These dust properties are dependent on the assumed dust model, but they shape the SED and so they can be seen as a different representation of the SED. We compare our method to energy balance SED fitting, a more traditional approach. To avoid a black box, we interpret the model in a variety of ways and investigate possible pitfalls.

In the next section, we present the data sets and algorithms that we used. Sect.~\ref{sec-sedfit} presents the SED fitting results, which are used as a reference, while Sect.~\ref{sec-ml-results} contains the machine learning results. In Sect. \ref{sec-interpretation} we interpret the model by considering when predictions fail, how the model performs on galaxies that do not resemble the training set, and by questioning what UV--MIR shape can be attributed to high or low FIR emission. The conclusions can be found in Sect. \ref{sec-conclusions}. Appendix~\ref{app-higherz} applies our model at higher redshifts ($z < 0.5$), and discusses K-correcting. An overview of different machine learning models and their performance is given in Appendix~\ref{app-table}.

Since the results of a machine learning algorithm can depend on its exact set-up, including the random initialisation, we make our code publicly available on GitHub under the codename \textit{FIREnet}\footnote{https://github.com/wdobbels/FIREnet} (far-infrared emission networks). All steps from raw input to final predictions can be followed using jupyter notebooks. An interactive version of some of this paper's figures are also available online\footnote{https://wdobbels.github.io/FIREnet}, which makes it possible to select a particular galaxy and view its SED.

\section{Data and methods}
\label{sec-datameth}

\begin{figure*}
	\centering
	\includegraphics[width=17cm]{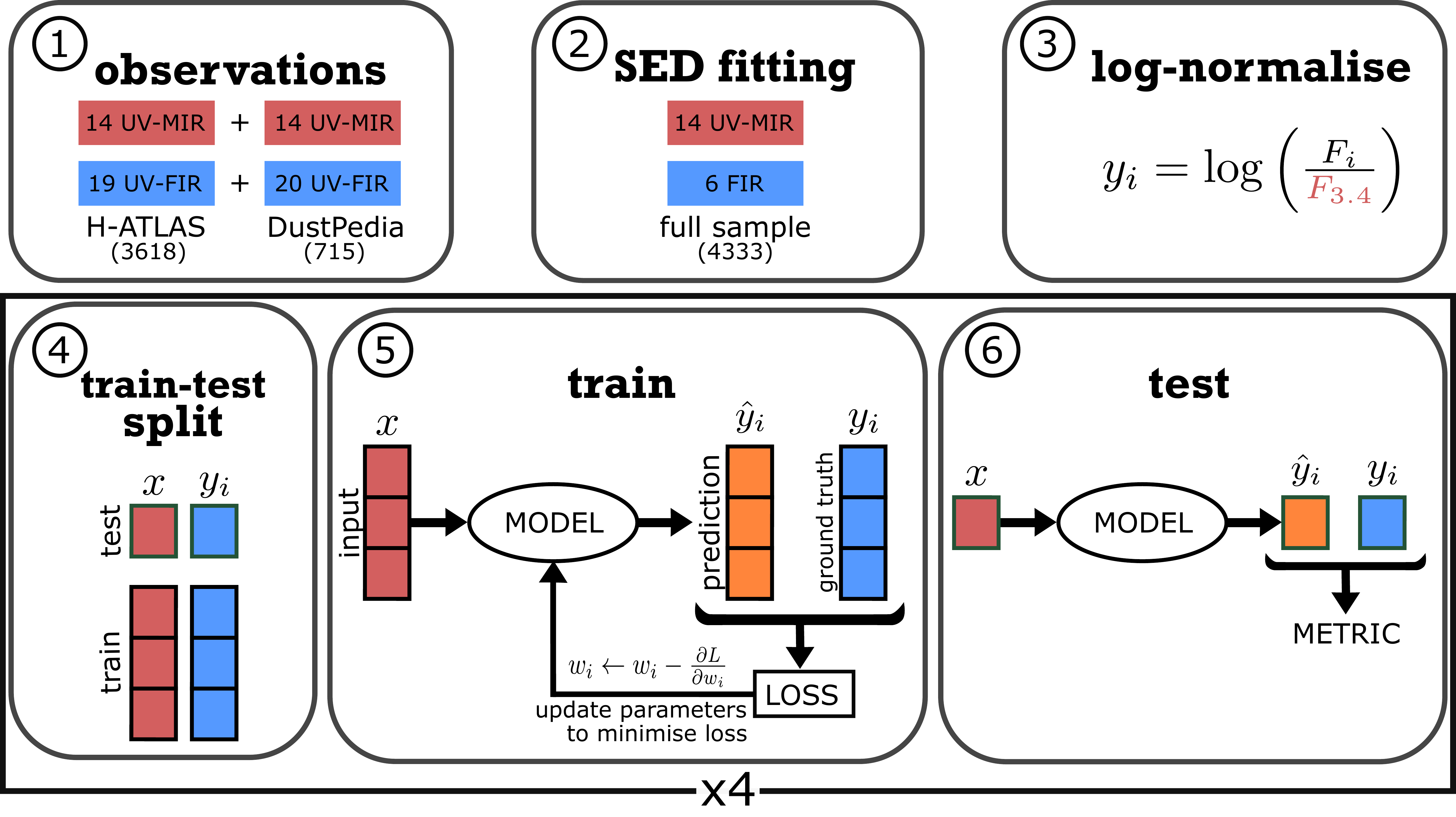}
	\caption{Diagram of the pipeline, split into six steps. The red boxes are used for input data, which do not make use of \textit{Herschel}. The blue boxes do require \textit{Herschel} observations and are used to derive the ground truth (i.e. prediction target). The orange boxes are model predictions. Steps 4 to 6 are repeated for the four folds, in order to use the full data set as a test set.}
	\label{fig-pipeline}
\end{figure*}

In this work, we apply supervised learning, in which a ML algorithm learns by example using a `training set'. An optimal mapping from features (i.e. input, the UV--MIR broadband fluxes) to target (i.e. output, the FIR broadband fluxes) is learned through optimisation. The goal of the algorithm is to have good predictions on unseen data, which we validate by setting apart a test set. Good predictions require a sufficiently complex (but not too complex) model, but also a large and reliable training set. The unseen data should be well represented by the examples that were used for training \citep{nalu}. In this section we discuss both the data set and the ML algorithm. Our complete pipeline is illustrated in Fig.~\ref{fig-pipeline}. 

\subsection{Data}
\label{ssec-data}

We need a data set that contains broadband fluxes from UV to FIR for a large number of galaxies. In order to have a larger, more complete sample, we combined two data sets. The first data set is the \textit{Herschel}-ATLAS (H-ATLAS) DR1 data \citep{hatlas-dr1}. It covers three equatorial fields observed by the GAMA survey \citep{gama}, amounting to a combined 161 sq deg. It includes only sources which have a detection above 4$\sigma$ in any of the SPIRE bands. The following bands are used:

\begin{itemize}
	\item GALEX: FUV and NUV \citep{galex}
	\item SDSS: $u$, $g$, $r$, $i$, $z$ \citep{sdss}
	\item VISTA: $J$, $H$, $K_s$ \citep{vista}
	\item WISE: \um{3.4}, \um{4.6}, \um{12}, \um{22} \citep{wise}
	\item \textit{Herschel} PACS: \um{100}, \um{160} \citep{pacs}
	\item \textit{Herschel} SPIRE: \um{250}, \um{350}, \um{500} \citep{spire}
\end{itemize}

These 19 broadbands sample the wavelength range between \um{0.15} and \um{500} relatively well. From the catalogue, we ensured that at least 5 UV--MIR fluxes were detected, as well as at least 3 FIR fluxes. We only used galaxies that have a spectroscopic redshift (of sufficient quality Z\_QUAL $\ge 3$) within 0.01 < z < 0.1. A local sample allows us to study galaxies over a large parameter space, whereas higher redshift galaxies must be luminous in the infrared to be detected by \textit{Herschel}. However, in Appendix~\ref{app-higherz} we demonstrate our method on a higher redshift sample ($0.1 < z < 0.5$), and show that a (spectroscopic) redshift is not required for our method to work. Since we make use of SED fitting (see Sect.~\ref{ssec-dataprep}), we also placed a threshold on the SED fit: we threw away the ten galaxies for which $\chi_r^2 > 10$. This leaves us with 3\,618 galaxies. 

The other data set is DustPedia \citep{dustpedia}: it is a UV--FIR data set containing nearby galaxies observed by \textit{Herschel}. After excluding the galaxies that have contamination flags due to nearby sources \citep{dustpedia-photometry} and that lack enough data points (minimum of 5 UV--MIR and 3 FIR), we ended up with 715 galaxies. The benefit of this extra data set is twofold. For one, it gives us extra data to train the ML. It also allows us to train on one data set and test on the other, giving an estimate of how the ML performs on different data (see Sect.~\ref{ssec-outside-set}). The two data sets are each reduced in their own self-consistent way, but the two pipelines are not the same. The DustPedia data set also has \um{70} data from PACS, and uses $J$, $H$, and $K_s$ filters from 2MASS \citep{2mass} instead of VISTA. Both data sets are corrected for Galactic dust extinction.

Besides the different pipelines, there are also intrinsic differences between the data sets. The H-ATLAS data set has a median redshift of 0.07, while for DustPedia this is only 0.004 (after correcting for proper motion). The DustPedia sample contains galaxies that are less luminous in the FIR, since the sample is more local ($z < 0.01$) and the constraints on FIR detections are less stringent (3 FIR bands need to be observed, but no detection requirements are imposed). DustPedia hence probes a more dynamic range at the low luminosity end, and is a good benchmark for local galaxies. However, it does not contain enough galaxies to properly train a ML pipeline (we need about 2~000 galaxies, see Appendix~\ref{app-higherz}). Even after limiting the redshift range, H-ATLAS allows us to have a sufficient number of galaxies for training. We combined the H-ATLAS and DustPedia galaxies into a single sample.

\subsection{Data preparation}
\label{ssec-dataprep}

We are dealing with a regression problem: predicting the FIR fluxes from the UV--MIR fluxes. The prediction targets are fluxes in the six \textit{Herschel} bands (\um{70} to \um{500}), while our input fluxes correspond to shorter wavelengths (from \um{0.15} to \um{22}). Throughout this work, all fluxes and luminosities are per frequency. 

One of the problems we had to deal with is unavailable fluxes. Most ML algorithms can not directly deal with missing data, and so a way of imputing these values is needed. In this case, the straightforward option is to interpolate. However, there are errors on the fluxes, and galaxies with missing data and large errors can lead to inaccurate interpolated fluxes. This is especially problematic for our target FIR fluxes. To avoid this, we created a physically motivated SED using the SED fitting tool CIGALE \citep{cigale, pcigale}. CIGALE fits a grid of model SEDs to the observed fluxes, and---using Bayesian inference---it calculates the posterior of the fluxes. While CIGALE is typically used to estimate physical properties, the latest version allows us to estimate the flux posteriors in exactly the same procedure. These Bayesian flux estimates were used as our ground truth, for all bands (not only for the missing values). We assume flat priors, and hence these Bayesian flux estimates are likelihood-weighted averages over all SED models. They provide good interpolations for fluxes that are missing, even when a series of bands from the same telescope are missing. For example, all of H-ATLAS lacks the \um{70} flux from PACS, but we will still use the Bayesian \um{70} estimate (although keeping in mind that it is less constrained than for the DustPedia galaxies). Moreover, fluxes that do not fall in line with neighbouring broadbands but contain large errors are pushed back towards a more reasonable value. The benefit of using the Bayesian fluxes (instead of the best fit models) is that we are not limited by the discreteness of the fitting parameter grid. Some example SEDs are shown in Fig.~\ref{fig-example-seds} (which is discussed in Sect.~\ref{ssec-goodbad}). For the DustPedia galaxies, we used the same $\sim 80$ million models as the DustPedia collaboration \citep{dustpedia-cigale}. For H-ATLAS, the redshift of the model grid was binned to 0.02, and we reduced the parameter grid to $\sim 38$ million models per redshift bin, by trimming values that always received a small likelihood.

Since the CIGALE fit makes use of all data (including FIR) to make the Bayesian fit, it is not fair to extract our input UV--MIR fluxes from this fit. After all, including the FIR fluxes could change the Bayesian MIR flux estimates in a way that makes it easier to predict the FIR fluxes. The FIR is not available when applying our algorithm `in the wild'. This is why each galaxy requires two CIGALE fits. The first has access to all fluxes, and is used to determine the FIR fluxes (our prediction target). The second only has access to the UV--MIR fluxes (up to \um{22}), and is used to determine the input features. This ensures that no data from the targets (FIR) leaks into the features (short wavelengths). We experimented with K-correcting the data, but (mostly due to additional uncertainties in WISE \um{22}) this did not improve the results (see Appendix~\ref{app-higherz}).

It helps if we can incorporate some prior knowledge into the algorithms. We start by normalising all fluxes to the \um{3.4} flux; $M_\star/L$ varies little in the \um{3.4} band and hence this band is a good tracer for the total stellar mass and luminosity \citep{mstar-34, meidt2014}. The \um{3.4} flux itself is instead converted to a luminosity. This means that all flux features are now distance independent. The distance can still be made available to the ML through an additional redshift feature, although we found that this was unnecessary (see Appendix~\ref{app-higherz} and the redshift model in Appendix~\ref{app-table}). The total luminosity is now only encoded into the \um{3.4} feature, and the other fluxes---technically colours with \um{3.4}---are now intensive (i.e. independent of the size of and distance to the galaxy). Next, we took the base-10 logarithm of all these features. Neural networks often struggle with data that spans multiple orders of magnitude, and taking the logarithm avoids this problem. In other words, the used features consist of all colours with respect to the \um{3.4} flux, as well as the logarithmised \um{3.4} luminosity. The target FIR fluxes are also divided by the Bayesian (UV--MIR) \um{3.4} flux and logarithmised (again resulting in colour-like variables). 

We used the same UV--FIR CIGALE fit to estimate dust properties in a Bayesian way: the dust luminosity $L_d$, dust mass $M_d$, and cold dust temperature $T_d$. These were derived from the THEMIS dust model \citep{themis}. The dust temperature only considers the cold dust (at the lowest radiation field intensity, see \citealt{dustpedia-cigale}), but the dust luminosity and mass include both the cold and warm dust. When predicting these properties, we first converted $L_d$ and $M_d$ into intensive properties (log($L_d/L_\star$) and log($M_d/M_\star$)), but convert them back to the extensive properties on our figures. The dust temperature required no further processing. These properties are model dependent, but can be seen as an underlying representation of the FIR shape; the dust temperature controls the colour of the FIR SED, while the dust luminosity controls the total luminosity in the FIR. 

While we need enough data to train the machine learning, we also want to assess its performance. For this reason, we kept apart a test set. The test set is only used after the model finished training, and provides an unbiased estimate of the model's performance on a similar data set (i.e. a sample with the same selection criteria). Instead of using a single test set, we used a 4-fold train-test split \citep{cross-validation}. This means that we randomly split our data set in four equally large parts, and train four models. Each model was trained on three parts and tested on the remaining part (25\% of the data). We can then use our full data set as an unbiased test set. Steps 4, 5, and 6 of the diagram in Fig.~\ref{fig-pipeline} are hence repeated four times for the different splits. If we need to apply our model to a galaxy from a different data set, we can either use one of the four models, or train a new model on the complete data set. For all figures involving ML predictions, we combine the four test splits, so all data points were unseen by their respective model.

\subsection{Flux predicting}
\label{ssec-fluxpred}

The machine learning process optimises a mapping from the input features to the targets, based on the training set. To predict the six target fluxes, we can either use six independent algorithms, or use a single algorithm (with a shared inner representation of the features) with six output values. After testing several algorithms, we decided to use neural networks as the main ML method. This produced among the best results, and allows for the six outputs with a shared inner representation. Other models, such as random forests and even linear models, are also competitive and can be used instead. The three dust properties were estimated with a similar but separate neural network, which has three output neurons. 

A neural network consists of a series of layers. Each of the layers calculates an output vector, where each value is a linear combination of the input vector. A non-linear function (the `activation function') is then applied to this vector, and the result acts as the input vector for the next layer. In other words, we apply a series of matrix multiplications to the input, each followed by a non-linear function, in order to calculate the final output vector (the prediction). Each of the layers can be seen as another representation of the data, and higher up layers consist of more abstract representations that are more related to the prediction targets. The last layer of our network did not use an activation function, so the predictions are a linear combination of the layer below. When making a prediction, the data flows forwards from the input layer (the 14 UV--MIR features) to the output layer (the 6 \textit{Herschel} targets).

Arbitrary functions can be approximated  \citep{nnet-universal-approximator}, and this optimisation is done by tuning the weights of all the matrix multiplications. Due to the large number of weights that need to be optimised, the only successful optimisation strategies (at the time of writing) are variations of gradient descent \citep{gradient-descent}. For this work we used the Adam optimiser \citep{adam}, and applied cosine learning rate annealing \citep{cosine-annealing}. The goal of the optimisation is to minimise a loss function (also known as `cost function') between the predicted outputs and the target. For the prediction of the FIR fluxes, we used a mean squared error (MSE) loss function (between the log-normalised fluxes as described in Sect.~\ref{ssec-dataprep}). We standardised the input and output of the neural networks to zero mean and unit standard deviation (fit on the training set). The data is unscaled for all results (including plots and metrics), and so this scaling can be seen as part of the inner workings of the ML model.

One of the problems that ML algorithms have to deal with is overfitting. This means that our network is too complex, and fits to the noise. The results on the training set are then very good, but predictions on the test set are poor. In other words: the network does not generalise. To avoid this, we use $L_2$ regularisation \citep[also known as weight decay;][]{weight-decay}, which adds a term to the loss function that pushes the weights closer to zero. 

Our model has three hyperparameters: the strength of the regularisation, the architecture of the network (number of layers and neurons in each layer), and the the non-linear activation function. These hyperparameters were not optimised through gradient descent, but by using 4-fold cross-validation. We only optimised these hyperparameters for the first train-test split, in order to have a uniform architecture over all four splits. The training set of this split was again divided in four: three parts are used for the actual training, while the final part is used for hyperparameter validation. This process is then repeated for the four cross-validation splits, and the parameters that performed best on the validation sets are used in our final model. This procedure resulted in a rectified linear unit (ReLU) activation function \citep[max(0, $x$);][]{relu}. For the architecture of the network, we use two hidden (i.e. inner) layers, both of which have 100 neurons, resulting in a total of 12~206 trainable weights. By today's standard, these are considered very shallow networks. A deep neural network, which is warranted for large data sets of complex problems, does not seem to be necessary.

Although the model is now fixed, we still set apart a validation set (25\% of the training data) in each train-test split. This set was used for early stopping (stop training after validation set performance convergences), and is also used for the uncertainty estimation (see next section). We built our framework using skorch\footnote{https://github.com/skorch-dev/skorch}, a scikit-learn compatible neural network library that wraps PyTorch \citep{pytorch}.

\subsection{Uncertainty estimation}
\label{ssec-methods-uncertainty-estimation}

While regression is a standard problem in machine learning, less attention has gone to predicting uncertainties. While the test set gives us a global estimate of how well the algorithm performs, we might have better predictions on some galaxies compared to others. In our case, we expect to have more accurate predictions when the errors on the fluxes are small, for example when the galaxy is bright.

In order to estimate the uncertainty, we used a second neural network: the uncertainty quantifier. We assume that our predictions (of the log-normalised fluxes) follow a Gaussian distribution around the ground truth (the Bayesian SED fit). This scatter around the true value represents the aleatoric uncertainty (irreducible with more data), which includes both the uncertainty on our ground truth (i.e. the prediction target) as well as the lack of information we have from the limited features: even if there were no observational errors and unlimited data, we expect that 14 UV--MIR broadband fluxes do not give enough information to perfectly predict the FIR fluxes. Our assumed error model can be written down as follows:

\begin{equation}
P(y \, | \, \textbf{x}, \mathcal{N}) = \frac{1}{\sqrt{2\pi \hat{V}(\textbf{x})}} \exp\left[ -\frac{\left( y - \hat{y}(\textbf{x}) \right)^2}{2 \hat{V}(\textbf{x})} \right].
\label{eq-error-model}
\end{equation}

The symbols with hats represent neural network predictions, with the flux $\hat{y}(\textbf{x})$ being predicted by the regression network and the variance $\hat{V}(\textbf{x})$ being predicted by the uncertainty quantifier. The prediction target is denoted by $y$: the Bayesian SED fitted log-normalised FIR flux. Equation \ref{eq-error-model} holds for each of the six \textit{Herschel} bands independently. The regression network minimises the MSE as described in the previous section, but for the uncertainty quantifier we minimised the negative log likelihood. For all points in the training set and all six \textit{Herschel} bands, we can write down the loss (disregarding the constant involving the $2\pi$) as follows:

	\begin{equation}
	\mathcal{L} = \sum_{\rm{train}} \sum_{\rm{FIR}} \frac{1}{2} \left[ \frac{\left( y - \hat{y}(\textbf{x}) \right)^2}{\hat{V}(\textbf{x})} + \ln\left( \hat{V}(\textbf{x}) \right) \right].
\label{eq-log-likelihood}
\end{equation}

This Gaussian likelihood loss function was proposed by \citet{nnet-uncertainty-gauss}. We have experimented with a more general approach (with a different parametrisation for the error model) from \citet{nnet-uncertainty-general}, but this did not noticeably improve the results and hence we opted for the simpler model. 

Since this neural network estimates uncertainty, it makes sense to add some extra features that describe the uncertainty on the input. In addition to the CIGALE fitted fluxes, we included two extra feature sets: the observed fluxes and their errors. Both of these feature sets were first normalised to their corresponding (UV--MIR) CIGALE fitted flux, after which we took the logarithm. This again results in intensive properties with a smaller range. Higher values denote more uncertainty on the input. For the bands where these values were not available (not observed or negative fluxes), we replaced the missing values by 6, which is about twice as high as the largest non-missing entry, and hence denotes very large uncertainty in that band. Instead of directly predicting $\hat{V}$, we predicted $1 / \hat{V}$. The final activation function of this network is the softplus function \citep{softplus}, which ensures positive values. It is possible to connect the regressor and uncertainty estimator (one uses the output from the other), but we found no clear improvements by doing so. In order to evaluate the uncertainty estimator, we define $\chi^2$:

\begin{equation}
\chi^2 = \frac{\left(y - \hat{y}(\textbf{x})\right)^2}{\hat{V}(\textbf{x})}.
\label{eq-chi}
\end{equation}

We assumed $\hat{y} \sim N(y, V)$, so the mean $\chi^2$ should be one. However, since the predictions on the training set are typically better than on the test set, our uncertainty estimator is biased at predicting uncertainties that are too small. To avoid this, we propose a correction factor based on the validation set. If we multiply each $\hat{V}$ with $\left< \chi^2_{\textrm{val}} \right>$, then by definition $ \left< \chi^2_{\textrm{val}} \right>$ becomes 1 after the correction. Since the uncertainty estimator is not trained on the validation set, this factor is typically larger than 1 (but smaller than 1.4 in our experiments). Multiplying with this factor hence compensates for the larger uncertainty on unseen samples. The test set is not used to calculate this correction factor, and hence remains unbiased.

Our approach captures the aleatoric uncertainty, which is both due to the uncertainties on the (input and output) fluxes, as well as the missing information in UV--MIR broadbands. It does not take into account epistemic uncertainty, which relates to a lack of data. When there is insufficient training data, different initialisations of our network will lead to different predictions. This class of uncertainty can be captured using ensembles of neural networks \citep{nnet-uncertainty-general}. However, we found that different neural networks lead to strongly correlated predictions, and hence the epistemic uncertainty is small compared to the aleatoric uncertainty. So for our fiducial model, we do not use ensembles and hence neglect the epistemic uncertainty. 

\begin{figure*}
	\centering
	\includegraphics[width=17cm]{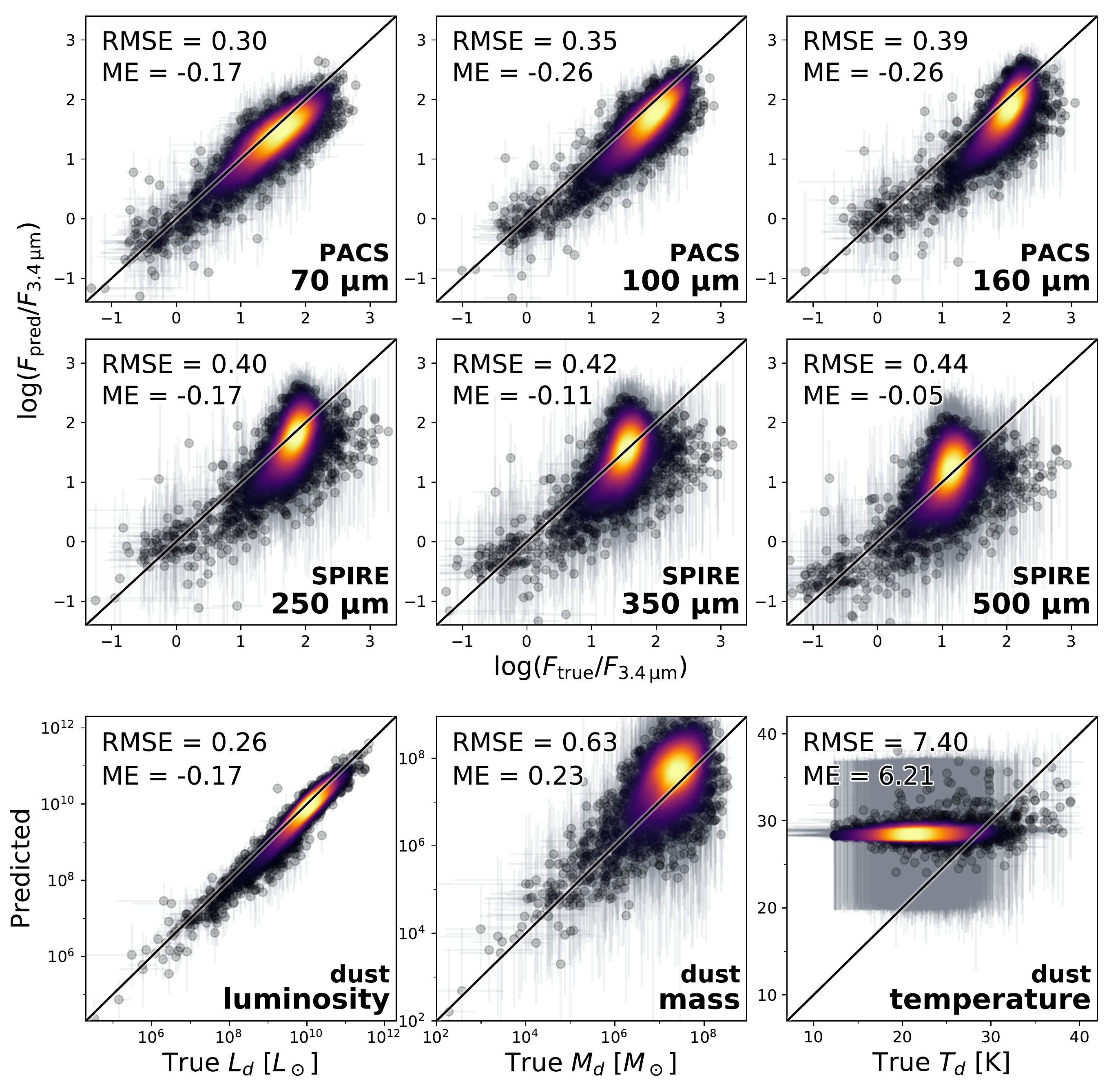}
	\caption{Predicted vs true values, for the sample of 4~333 galaxies (H-ATLAS + DustPedia). The predictions are derived from Bayesian SED modelling to the UV--MIR broadband fluxes. The ground truth values are also derived from a Bayesian SED fit, but instead make use of the full UV--FIR broadband fluxes. The error bars are derived from the likelihood-weighted standard deviations reported by the CIGALE fits. The colour of the points reflects a 2D kernel density estimate, with brighter colours corresponding to denser regions. The top two rows show the log-normalised fluxes ($F_{\um{3.4}}$ is the Bayesian estimate of the WISE \um{3.4} band from the UV--MIR SED fit). The bottom row shows intrinsic dust properties. Each panel lists the RMSE and ME (ME = $\hat{y} - y$). For the dust luminosity and dust mass, these are calculated in log-space (dex), but for the dust temperature they are not.}
	\label{fig-truevspred-cigale}
\end{figure*}

\section{SED fitting predictions}
\label{sec-sedfit}

Before we show the FIR predictions from the machine learning, we first introduce a more traditional SED fitting approach. In order to make a fair comparison, both methods use the same data set. As described in Sect.~\ref{sec-datameth}, this data set is a combination of H-ATLAS (3\,618 galaxies) and DustPedia (715 galaxies), totalling 4\,333 galaxies. We are interested in predicting the FIR flux in six \textit{Herschel} broadbands, as well as the dust luminosity, mass, and temperature.

The UV--MIR CIGALE fits introduced in Sect.~\ref{ssec-dataprep} can be used to extract FIR predictions. This is similar to how the ground truth values are extracted, but without constraining the FIR broadband fluxes with observations. The resulting predictions (of the log-normalised fluxes) for each of the six \textit{Herschel} bands are shown in Fig.~\ref{fig-truevspred-cigale}, in the top two rows. The RMSE is also shown for each of the bands. The combined RMSE is 0.38 dex. At shorter FIR wavelengths, the predictions match the ground truth well, although they are slightly biased towards underpredictions. The predictions degrade towards longer wavelengths. Especially noticeable is the increase in bias: the trend becomes curved, with large underpredictions for intermediate values. The benefit of this Bayesian SED fitting approach is that it directly estimates the uncertainty (even a full posterior probability density function if needed). We see indeed that the uncertainties for the predictions (fit not constrained by \textit{Herschel}) are quite a bit larger than for the ground truth (fit constrained by \textit{Herschel}). However, the size of these uncertainties depends on the width of the assumed priors.

The prediction of the dust properties is shown in the bottom row of Fig.~\ref{fig-truevspred-cigale}. The dust luminosity ($L_d$) can be estimated quite well (RMSE = 0.26~dex). This $L_d$ is directly constrained by the (assumed) energy balance: it is the same as the absorbed stellar luminosity (which relies on a good estimate of the unattenuated stellar spectrum). However, the success of these predictions is not due to the UV--NIR energy budget. When leaving out the WISE bands (in which the dust absorption is negligible), the $L_d$ estimate considerably degrades. The RMSE on $\log (L_d)$ increases to 0.40 dex (overall RMSE over the six FIR bands becomes 0.44 dex). For the figure that excludes the four WISE bands, refer to Appendix~\ref{app-figures}, Fig.~\ref{fig-truevspred-cigale-nowise}. The estimation of $L_d$ only works well because the inclusion of the MIR (\um{12} and \um{22}, originating from warm dust) constrains the FIR. Relying on energy balance alone (when the MIR is not available) requires better priors on the unattenuated UV--MIR spectrum for this data set. 

The dust mass is harder to predict, with a RMSE of 0.63~dex. The dust mass correlates with the dust luminosity which is predicted reasonably well, but other unknown dust properties (cold dust temperature, cold dust fraction and small hydrocarbon fraction) increase the scatter. The cold dust temperature $T_d$ is essentially unconstrained, and so the predictions are close to a constant value. This is not really surprising: the WISE \um{12} and \um{22} bands may give a good hint of the dust luminosity and mass, but do not help constrain the cold dust temperature. It is remarkable that the predictions cluster around a mean temperature of 28.7~K, while the average true temperature is only 22.4~K. One might think that this is due to the choice of the CIGALE grid (the prior), but the grid in $U_{\textrm{min}}$ (diffuse interstellar radiation field; ISRF) corresponds to an average (median) temperature of 23.1~K (21.7~K). The dust emission model assumes a \citet{draineli2007} ISRF: a fraction 1 - $\gamma$ is heated by $U_{\textrm{min}}$ (the cold dust), while a fraction $\gamma$ is heated by a power-law distribution from $U_{\textrm{min}} < U < U_{\textrm{max}}$. We found that in the absence of FIR constraints, the UV--MIR fits will generally call for warmer dust in order to help fit the WISE bands. This leads to a larger $U_{\textrm{min}}$, increasing the temperature for both the warm and cold dust which leads to the overestimation of $T_d$. In any case, the uncertainties on these predictions are large, and properly reflect the fact that the cold dust temperature can not be constrained from the UV--MIR with an SED fitting approach.

Up to this point, we have used RMSE as a performance metric. It is however also worth quantifying biases. A quantity that reflects this is the mean error (ME = $\hat{y} - y$), which is defined to be positive for global overpredictions and negative for underpredictions. All of the predicted \textit{Herschel} bands are underpredicted on average. PACS 100 and \um{160} have the largest biases, both averaging a misprediction of -0.26 dex. The result is that the total dust luminosity is also underpredicted, with an average error of -0.17 dex. When excluding the WISE bands in the fitting, the mean error of $L_d$ is below 0.01 (even though the RMSE is much larger; see Fig.~\ref{fig-truevspred-cigale-nowise}). In both cases (with and without WISE), the fitted SEDs seem to have a too low FIR-to-MIR ratio on average: too much warm dust, not enough cold dust. Without WISE, we overestimate the MIR and underestimate the FIR, leading to an unbiased $L_d$; with WISE, the MIR is constrained but we underestimate the FIR, and so the total dust luminosity is underestimated as well.

The dust temperature has a mean error of 6.2~K (overpredicted). Even though this would lead to a lower cold dust mass (a higher temperature leads to a higher dust luminosity at fixed mass, and so less mass is needed at fixed luminosity), we find that the dust mass is on average overestimated (mean error of 0.23 dex). This can again be attributed to a large portion of dust being present in a warm component. Once the \textit{Herschel} bands are included, the unrealistically large warm dust mass is ruled out.

\begin{figure*}
	\centering
	\includegraphics[width=17cm]{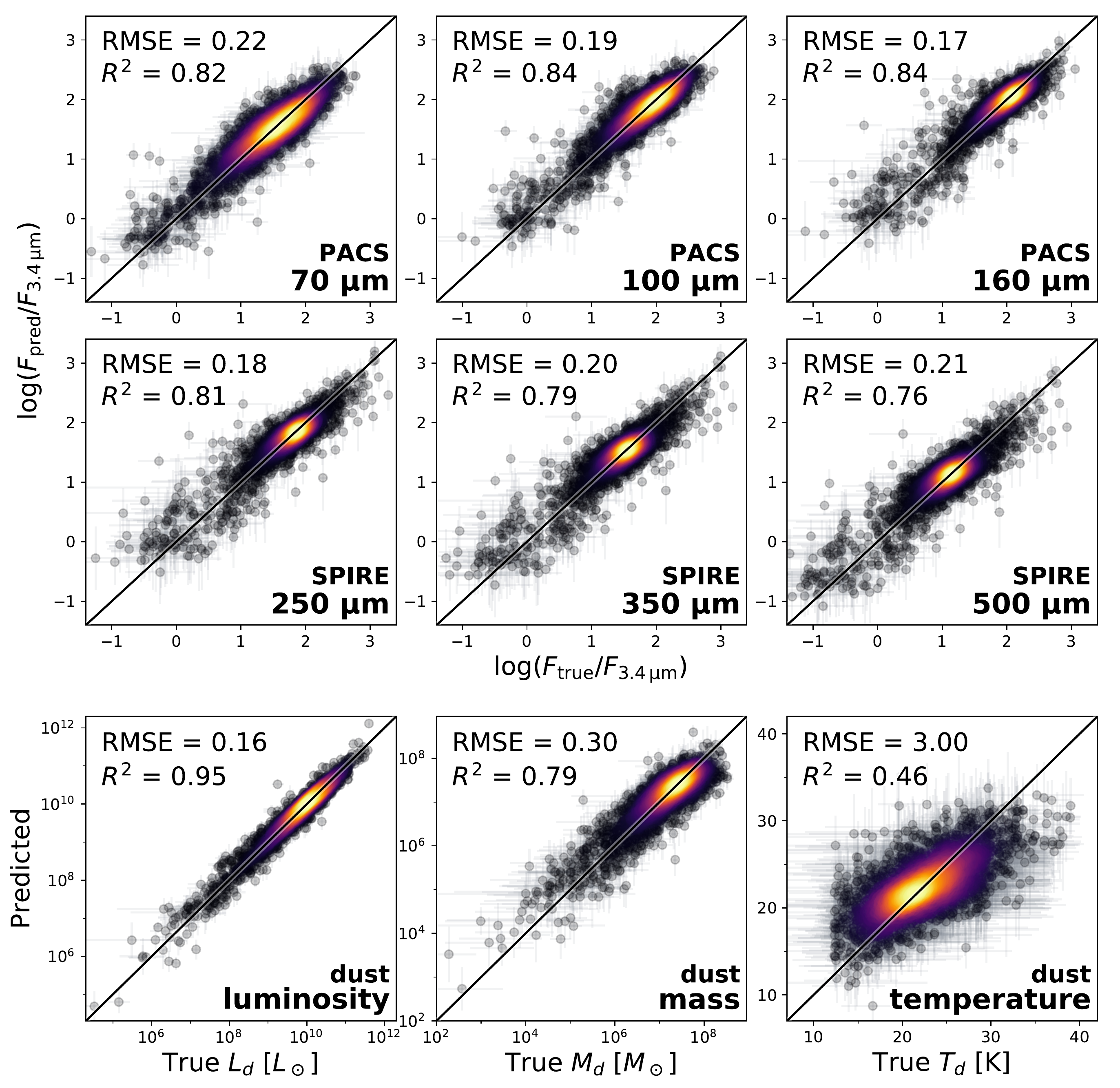}
	\caption{Similar to Fig.~\ref{fig-truevspred-cigale}, but for the neural network prediction. Instead of ME, which was always very small, we show $R^2$ (Eq. \ref{eq-r2}).}
	\label{fig-truevspred-nnet}
\end{figure*}

\section{Machine learning predictions}
\label{sec-ml-results}

The main limitation of the SED fitting approach is that the dust emission is modelled independently from the dust absorption, save for the energy balance assumption. Given enough data, machine learning can learn more complex patterns. For example, we know that although elliptical galaxies tend to have much lower star formation rates (SFR), their cold dust temperature tends to be higher \citep[e.g.][]{dustpedia-cigale}. The SFR can be estimated from the shorter wavelengths, and this correlation can then indirectly be used by a machine learning approach to estimate $T_d$.

\subsection{General results}
\label{ssec-ml-general}

The neural network predictions are shown in Fig.~\ref{fig-truevspred-nnet}, which has an interactive equivalent on the GitHub page. The global RMSE of the six \textit{Herschel} bands is 0.19~dex, about halve that of the SED fitting approach. The best predictions (according to a RMSE metric) are for PACS \um{160} (RMSE = 0.17 dex), with the predictions being slightly worse (up to 0.22 dex) for both longer and shorter wavelengths. It is worth noting that at longer wavelengths, our target is more clustered around an average value. This makes the predictions easier, and increases the performance of a baseline predictor, which always predicts the mean $\bar{y}$. This is taken into account by an $R^2$-metric \citep{r2score, r2score-2}, which is a renormalised MSE metric that is 0 for a baseline predictor and 1 for perfect predictions (and thus unlike MSE, higher is better). It is defined as follows:

\begin{equation}
	R^2 = 1 - \frac{\sum_i \left(y_i - \hat{y}_i \right)^2}{\sum_i \left(y_i - \bar{y} \right)^2},
\label{eq-r2}
\end{equation}

\noindent where $\bar{y}$ is the mean of the ground truth values, and the summation index $i$ runs over all samples in the particular data set (usually the test set). With this metric, the predictions for PACS \um{70} ($R^2 =0.82$) are considered better than for SPIRE \um{500} ($R^2 = 0.77$), although the maximum still occurs for PACS \um{160} ($R^2 = 0.85$).

The predictions on the H-ATLAS galaxies (RMSE = 0.17) seem to be better than for DustPedia (RMSE = 0.29). DustPedia contains galaxies with a lower FIR luminosity, and these have larger uncertainties on the ground truth. The UV--MIR CIGALE predictions have a similar but smaller effect (RMSE = 0.38 for H-ATLAS, RMSE = 0.42 for DustPedia).

The improvements over an SED fitting approach are particularly clear when predicting the dust properties. This is shown in the bottom row of Fig.~\ref{fig-truevspred-nnet}. While we have shown that UV--MIR SED fitting works reasonably well to predict the dust luminosity, we can do better. The RMSE is 0.16 dex, compared to 0.26 dex for the SED fitting. This means that our method could be used in UV--MIR SED fitting to improve the estimation of the total absorbed energy. This can be readily implemented with CIGALE: first a normal UV--MIR run is performed, in order to get the Bayesian estimate of each of the 14 input bands. Then, the neural network estimates $L_d$. These values can then be used in a second CIGALE run to constrain the energy budget. This is similar to the SED+LIR fitting introduced by \citet{gswlc-2}. They use \citet{chary-elbaz-templates} templates to translate WISE \um{22} into $L_d$, recalibrate to their sample, and report a RMSE of $\sim 0.1$~dex (on a different sample and without the use of a test set).

The improvement in dust mass prediction is even larger: a RMSE of 0.30~dex compared to 0.63~dex for the SED fitting. This is not surprising, since estimating $M_d$ with UV--MIR SED fitting relies on $L_d$; no other dust properties are directly constrained. While dust temperature is clearly a more difficult property to predict (even with a machine learning approach), the predictions are definitely better than the baseline, since $R^2 = 0.46$ (and RMSE = 3.0 K). 

\subsection{Prediction bias and regression towards the mean}
\label{ssec-pred-bias}

There are multiple ways to quantify a particular bias. The mean error for all bands is around 0.01, and hence negligible compared to the variance. The predictor is trained in a way such that each prediction is unbiased ($y - \hat{y}$ averages to zero). If a particular bin in $\hat{y}$ would on average be an underestimation, the neural network would adapt by increasing the predictions for that bin, hence removing that bias. As long as the test set is well represented by the training set, a machine learning approach is inherently unbiased.

Although our predictor is unbiased, one might notice that we tend to underestimate the largest $y$, while we overestimate the smallest $y$ (especially noticeable for the dust temperature). This is not a bias, but the consequence of regression towards the mean \citep{regression-mean}. The ground truth can be split into two parts: one part depends on the features, the second part is independent of our features. This second part consists of both missing information (e.g. star-dust geometry, grain size distribution) and the intrinsic uncertainties on the ground truth (related to the observational errors). Since this part by definition does not correlate with the features, our model can consider it random noise. When we take a galaxy with a large $y$, both parts will tend to contribute positively: the part of $y$ that correlates with the features, as well as the `random noise'. Our model can only determine the part that correlates with the features, and hence its prediction is closer to the mean. An extreme example is the baseline predictor, which is unbiased but always underpredicts large $y$ and overpredicts small $y$. Regression towards the mean happens for all bivariate distributions that correlate imperfectly. Regarding Fig.~\ref{fig-truevspred-nnet}, fitting a line by horizontal least-squares results in a slope of 1.0 (unbiased); fitting a line by vertical least-squares results in a slope smaller than 1. This is an expected consequence of the missing information, and should not be corrected for.

\subsection{Alternative models}
\label{ssec-alternative-models}

As described in Sect.~\ref{ssec-fluxpred}, the network's hyperparameters were optimised using 4-fold cross-validation. We also tried various machine learning algorithms besides neural networks, such as linear models \citep{ridge}, random forests \citep{random-forest}, gradient boosting \citep{gradient-boosting} and extremely randomised trees \citep{extremely-randomized-trees}. After optimising the hyperparameters of these algorithms, the results are very similar to those of the neural network. For example, a random forest has RMSE = 0.21~dex, while a linear model has RMSE = 0.24~dex (mostly the longest wavelengths are worse than neural networks). For an the RMSE of some of these models in the six Herschel bands, see Table~\ref{tab-overview}. We also tried an ensemble of different models (such as neural networks with different architectures), and found that these barely improve the results, so we decided to stick with the simpler model of a single neural network. 

While neural networks might be associated with long computations on large GPU clusters, this is not the case for our models. Training our fiducial model on one of the train-test splits takes around one minute on a modern dual-core i5 laptop, while predicting on all 4~333 samples takes about 100 ms. The largest constraint is running CIGALE to extract the Bayesian fluxes, which takes a few hours on a 20-core cluster.

\subsection{Machine learning uncertainty estimation}
\label{ssec-ml-uncertainty-results}

\begin{figure*}
	\sidecaption
	\includegraphics[width=12cm]{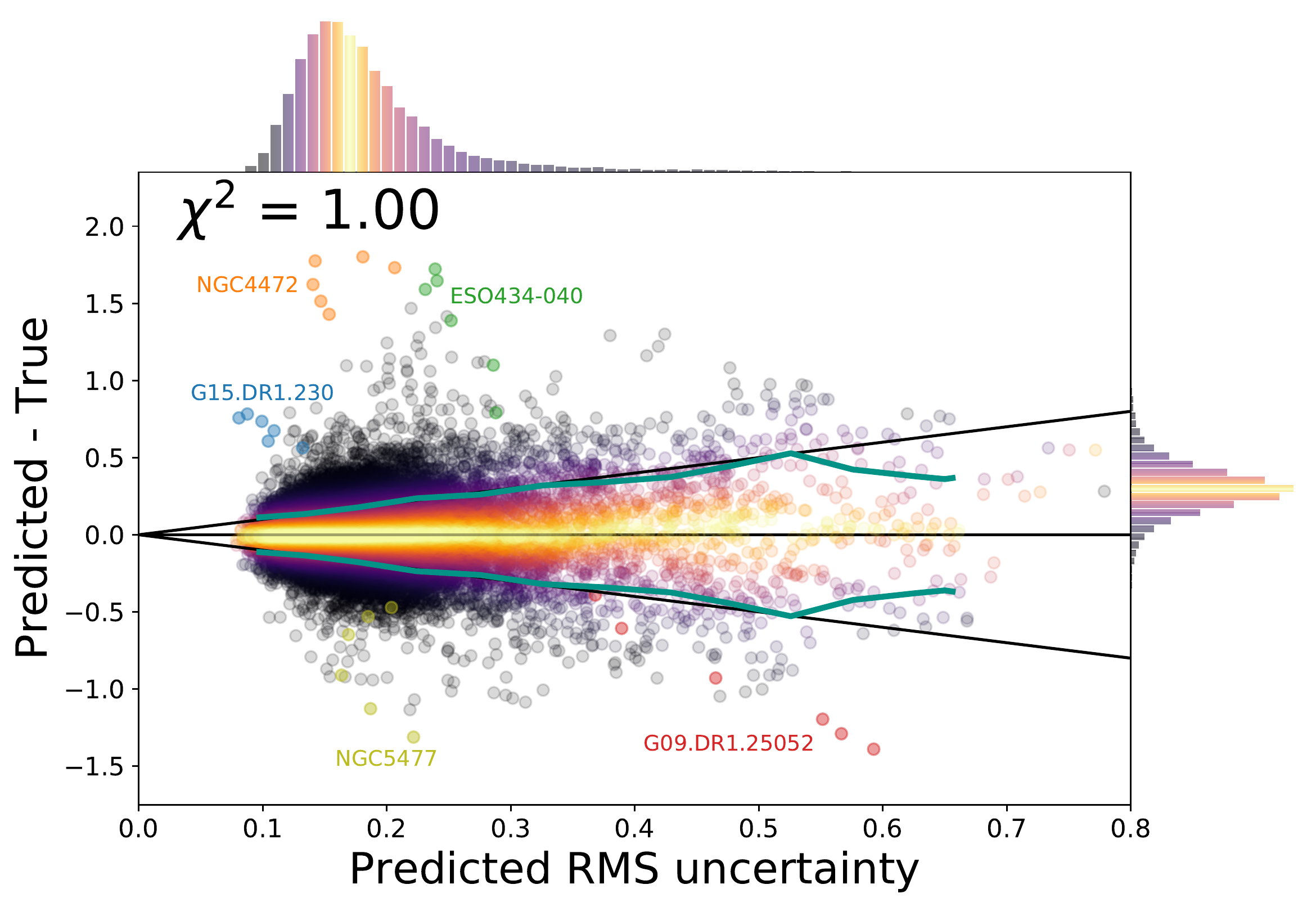}
	\caption{Misprediction error ($\hat{y} - y$) as a function of the predicted uncertainty. The values for all \textit{Herschel} bands are plotted on this figure, so each galaxy in the sample has six data points on this plot. The figure shows the (combined) test set of the 4-fold train-test split. The cyan lines show the RMS of the misprediction error ($\hat{y} - y$), and should be close to the black one-to-one lines if the predicted uncertainty is accurate. The colour shows the percentile in that bin, with the brightest values being used for the median. Two histograms show the marginal distributions of the two coordinates. A few outliers are indicated, using the same colour for all six bands of a single galaxy. An interactive version of this plot---where a galaxy can be selected to inspect its SED---can be found on the GitHub page.}
	\label{fig-uncertainty-nnet}
\end{figure*}

The uncertainty quantifier (described in Sect.~\ref{ssec-methods-uncertainty-estimation}) estimates the uncertainty on each prediction. We assume that each prediction follows a Gaussian distribution around the true value (see Eq.~\ref{eq-error-model}), and predicted the variance $\hat{V}$. In Fig.~\ref{fig-truevspred-nnet}, the predicted standard deviation (square root of the predicted variance) is shown as error bars. However, this figure does not allow us to assess the quality of our uncertainty estimation. 

In Fig.~\ref{fig-uncertainty-nnet}, we show the misprediction error $\hat{y} - y$ as a function of the predicted standard deviation. If this predicted uncertainty is valuable, it should match the RMS of the misprediction error, which is shown as the cyan line. We see indeed a good match between this cyan line and the black one-to-one lines. If we quantify this using a $\chi^2$ metric (Eq.~\ref{eq-chi}), we find that the mean $\chi^2$ of the sample is 1.00. In each of the bins of the predicted uncertainty, we also find that it is very close to 1. When considering the individual bands, the mean $\chi^2$ ranges from 0.97 to 1.05. The assumption of Gaussian errors works well, and we can accurately estimate its standard deviation. 

There are some outliers, which we discuss more in depth in Sect.~\ref{ssec-goodbad}. Figure~\ref{fig-uncertainty-nnet} shows all six predicted bands on the same figure. The errors made on different bands are not independent, and a single galaxy with bad predictions leads to multiple outliers on this figure. In the online interactive version of Fig.~\ref{fig-uncertainty-nnet}, the SED of these outliers can be inspected. The factor that corrects for the increased uncertainty on the test set (the uncorrected $\left< \chi^2_{\textrm{val}} \right>$) is 1.14 on average, but ranges (between bands and between train-test splits) from 1.07 to 1.24.

\section{Interpretation and discussion}
\label{sec-interpretation}

\subsection{Performance and sources of misprediction}
\label{ssec-goodbad}

\begin{figure*}
	\centering
	\includegraphics[width=17cm]{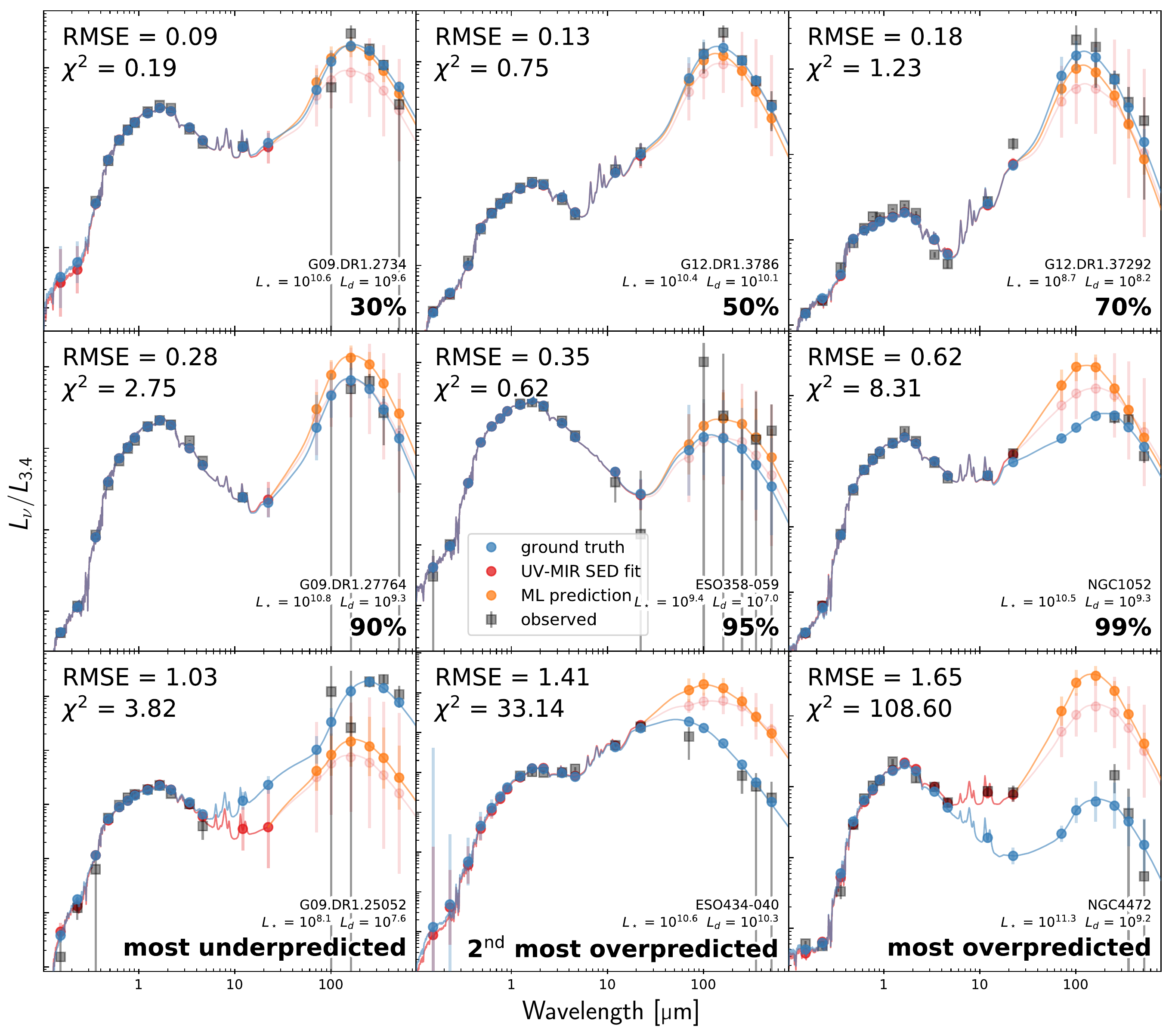}
	\caption{SEDs of nine galaxies, each shown in a different panel. The galaxies were sorted by RMSE, and we show galaxies at different percentiles (shown at the bottom right). The black markers are the observed fluxes, the blue markers are the ground truth (UV--FIR SED fit), the red markers are the UV--MIR SED fit, while the orange markers show the neural network predictions. The lines are a variation on the the UV--FIR best model SED, rescaled at each point to go through the respective broadbands; they are only meant to guide the eye. The luminosity is normalised to the \um{3.4} luminosity, but $L_\star$ and $L_d$ are given for each galaxy (in solar units). }
	\label{fig-example-seds}
\end{figure*}

Our goal is to predict the FIR part of a galaxy's SED, so some of these SEDs are shown in Fig.~\ref{fig-example-seds}. To avoid a random selection of galaxies, we ranked the galaxies according to their total RMSE, and show the galaxies at certain percentiles. Even the 90\textsuperscript{th} percentile has an RMSE of only 0.28~dex, and overall the $\chi^2$ (the mean $\chi^2$ across the FIR bands) is reasonably low: worse predictions are indeed typically compensated by larger predicted uncertainties. 

The only exception is when considering the worst samples. The RMSE can exceed 1~dex, and this is not always compensated by an adequately large uncertainty. We see that for the case of the most underpredicted galaxy (bottom left), the WISE~\um{12} and \um{22} bands were missing. The UV--MIR CIGALE fit tries to estimate these (properly having large uncertainties). When these MIR estimates are far from the ground truth (UV--FIR SED fit estimates), the neural network can be misled.

In the case of the most overpredicted sample (NGC 4472, also known as M49, bottom right), the ground truth WISE \um{12} and \um{22} bands are almost an order of magnitude below the observed bands. In this case, the SED modelling templates can not fit the high MIR emission together with the low FIR emission. Only a few galaxies have SPIRE luminosities lower than the WISE \um{3.4} luminosity (in per-frequency units), and combined with the high MIR luminosity there are not enough training samples for a good machine learning prediction. Moreover, from the \textit{Herschel} bands, only SPIRE \um{250} is detected at the $1\sigma$ level.

   \begin{figure}
	\centering
	\includegraphics[width=\hsize]{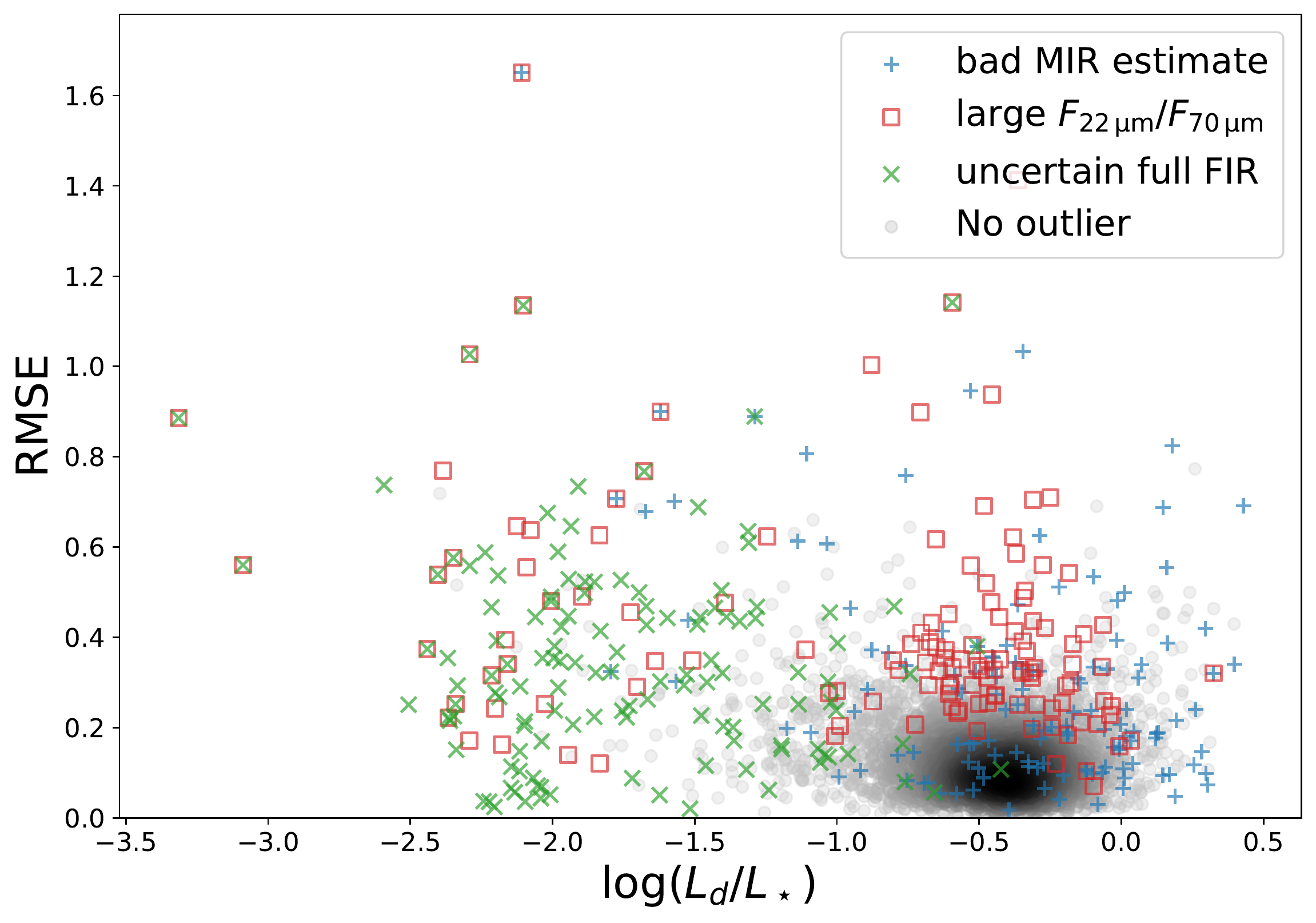}
	\caption{RMS prediction error (over the six FIR bands), as a function of log($L_d / L_\star$). Three sources of error use different marks and colours, and we show the 3\% worst samples of each category. Bad MIR estimate (blue) means that the \um{12} and \um{22} bands, estimated from the UV--MIR SED fit, do not agree with the same bands estimated by the UV--FIR SED fit (often because these lack observational constraints). Large $F_{\um{22}} / F_{\um{70}}$ (red) uses a Bayesian \um{22} from the UV--MIR SED fit but \um{70} from the UV--FIR SED fit. The uncertain full FIR (green) denotes galaxies with a large Bayesian (UV--FIR) relative uncertainty on the \textit{Herschel} bands.  Grey dots are not in the worst 3\% for any of the three categories. }
	\label{fig-errorsources}
\end{figure}

From the bottom row of Fig.~\ref{fig-example-seds}, we note two possible caveats of our predictor. The first is when the WISE \um{12} and WISE \um{22} bands are improperly estimated by the UV--MIR SED fitting. This is usually due to these two bands not being observed. Often, the UV--MIR SED fit still does a good job of estimating these two bands. However, as we will see in Sect~\ref{ssec-nowise} and Sect~\ref{ssec-data-interpretation}, these bands are important to the predictor; when they are poorly estimated, the predictions tend to be off. A second, related, caveat is when the MIR emission is unusually high for the FIR. This MIR emission traces a warm dust component, heated by an intense radiation field (e.g. an active galactic nucleus or intense star formation). The neural network can not properly detect this class of galaxies, either due to not having enough training samples of this kind, or because this class of galaxies can not easily be identified from UV--MIR SED alone. Of course, a third case in which we expect to be further from the ground truth is when there is a large uncertainty on the ground truth. In other words, when the UV--FIR SED fit has large Bayesian FIR uncertainties (due to large observational FIR uncertainties), our predictions might be far from this ground truth.

The effect of these three sources of additional uncertainty are shown in Fig.~\ref{fig-errorsources}, where we show the worst 3\% for each uncertainty category in blue, red, and green. The total FIR RMSE is shown as a function of log($L_d / L_\star$). Clearly, most outliers (e.g. RMSE > 0.5~dex) can be attributed to one or more of these error sources. While 3.8\% of the complete sample has a RMSE larger than 0.4~dex, this fraction reduces to 3.2\% when leaving out the bad MIR estimates (3\% worst, shown in blue), to 2.6\% when leaving out the galaxies with a large $F_{\um{22}}/F_{\um{70}}$, and to 2.8\% when leaving out the galaxies with an uncertain full FIR. However, when leaving out all galaxies that match one of these criteria, only 1.4\% of the sample has an RMSE larger than 0.4~dex, and none have an RMSE larger than 0.8~dex. Unfortunately, all three of the error sources require the full SED fit (and hence the FIR observations): these are diagnostic tools but can not be used to prevent a bad prediction (or prevent it from being used). However, when WISE \um{12} and WISE \um{22} observations are missing, it is better to use a retrained predictor that does not make use of these two bands (see Sect.~\ref{ssec-nowise}). The worst ground truth FIR fluxes are for galaxies with a low dust-to-stellar luminosity: their low FIR fluxes make them harder to detect. The other two error sources are situated at higher dust luminosities.

\subsection{Predicting on a different data set}
\label{ssec-outside-set}

\begin{figure*}
	\centering
	\includegraphics[width=17cm]{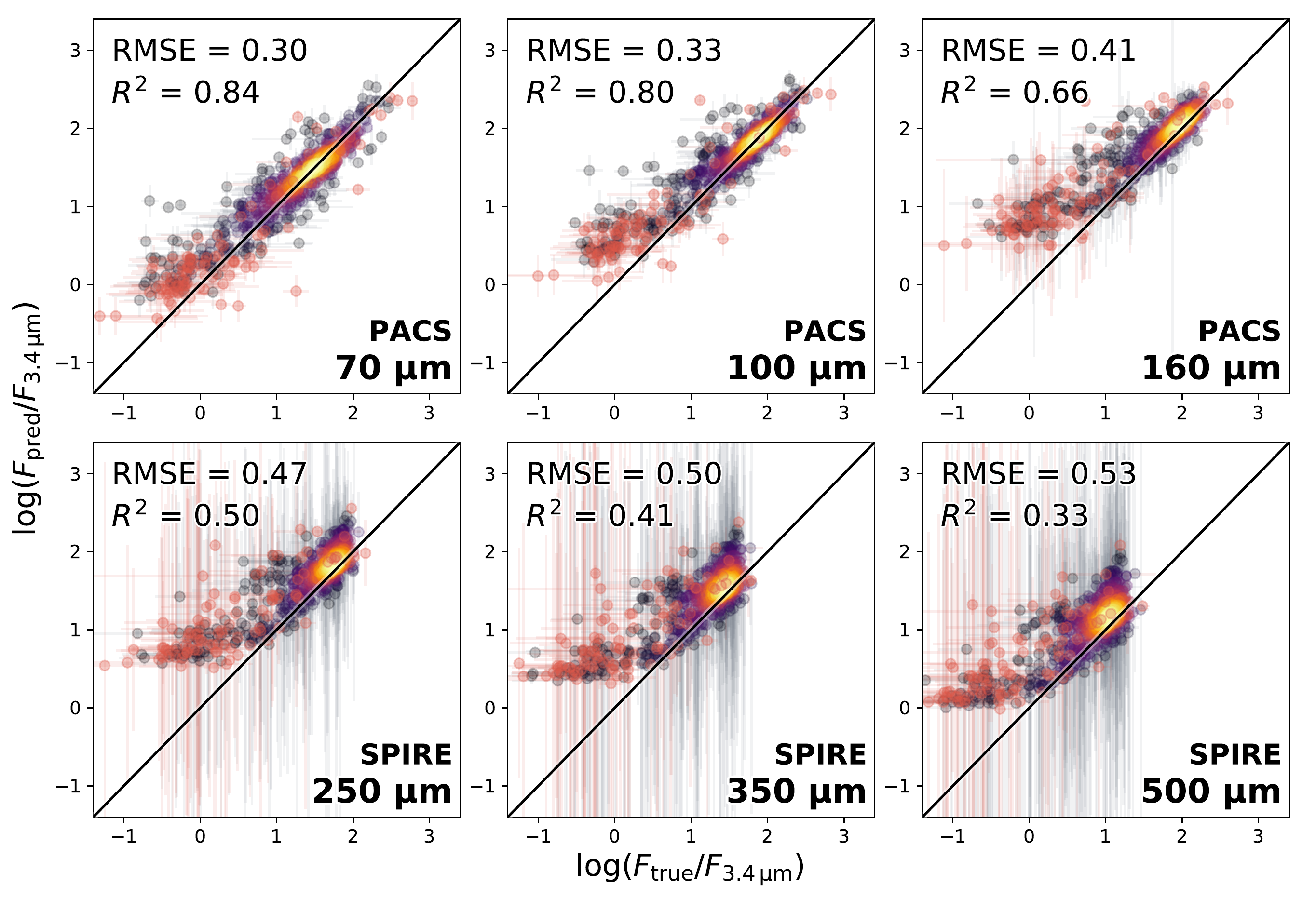}
	\caption{Similar to Fig.~\ref{fig-truevspred-nnet}, but for a model trained on H-ATLAS and tested on DustPedia (shown here). Only the log-normalised fluxes are shown. The red points are galaxies that do not resemble the training set (see main text), although the two metrics do include these flagged galaxies.}
	\label{fig-truevspred-nnet-hax}
\end{figure*}

One of the main drawbacks of machine learning is that it is often unsuccessful when used on samples that do not resemble the training set. The mapping from features to target is based on the training set, and within this feature space the mapping is usually well constrained. However, when a new data point is far from the training samples (in feature space), the prediction relies on an extrapolation, which can be inaccurate \citep{nalu}. 

In Fig.~\ref{fig-truevspred-nnet-hax}, we show a neural network that was trained on H-ATLAS and tested on DustPedia. The predictions $\hat{y}$ on the DustPedia set are shown as a function of the true value $y$. Unsurprisingly, the results are worse than within-sample testing (in which case the test set follows the same distribution as the training set). However, overall we still find a reasonably good prediction, with a total RMSE of 0.43 dex and positive $R^2$ in all bands. This is very similar to the performance of the SED fitting approach (Sect.~\ref{sec-sedfit}), which has a total RMSE of 0.42 dex on the DustPedia data set. We found small variations between different initialisations of the neural network (of the order of 0.02 dex), reflecting the larger uncertainty due to extrapolating. Hence, instead of using a single neural network, we used the average prediction of five neural networks (same architecture but different random initialisation) in this section. This ensembling did not really improve the results, but leads to more robustness and hence better reproducibility.

One modification to the architecture of Sect.~\ref{sec-ml-results} was necessary: the removal of the WISE \um{3.4} feature. As described in Sect.~\ref{ssec-dataprep}, the WISE \um{3.4} flux was used to normalise the fluxes in the other bands, and the log of this \um{3.4} luminosity was then an independent feature which served as a measure of the total luminosity. The DustPedia data set has more nearby galaxies than H-ATLAS, and contains galaxies with a lower (\um{3.4}) luminosity than H-ATLAS. The original network makes use of this \um{3.4} luminosity and extrapolates. We found that this leads to large overpredictions, leading to negative $R^2$ (worse prediction than baseline) at the longest two wavelengths. A network that does not make use of the \um{3.4} feature does not suffer from these extreme outliers, and has overall better results. For our fiducial mixed sample model (Sect.~\ref{sec-ml-results}), the \um{3.4} feature is still beneficial, especially at longer wavelengths: without it the SPIRE \um{500} RMSE is 0.28~dex (up from 0.21~dex).

After removing the WISE \um{3.4} band, one clear bias remains: at longer wavelengths, the predictions plateau at a lower limit, never going below a certain value. This leads to overpredictions where the actual value $y$ is below this lower limit. The reason for this is clear: DustPedia has less strict FIR detection criteria than H-ATLAS, and hence contains galaxies with a lower \textit{Herschel} to \um{3.4} flux ratio. Since the neural network has not seen galaxies with such low FIR emission, it is biased to overestimate the FIR of these galaxies. We have experimented with Neural Arithmetic Logic Units \citep[NALU;][]{nalu}, but did not find a satisfying improvement without fine-tuning the network's architecture for the test set. A linear model was also inferior to our neural network. Extrapolation for complex tasks like these is hard, if not impossible.

Fortunately, we do know when our predictions can be considered an extrapolation. If a new data point does not have any training set neighbours in feature space, the predictor is not well constrained and the predictions are uncertain. In Fig.~\ref{fig-truevspred-nnet-hax}, we highlighted 100 galaxies (in red) where the Euclidean feature space distance to its closest ten training set neighbours was largest. These galaxies fall mostly in the region of low FIR fluxes, where the bias of our predictor is largest. Even if we did not have \textit{Herschel} observations of the test set, we know that if the UV--MIR does not resemble the training set, the prediction is not reliable. The overall test RMSE when excluding these points is 0.36 dex (0.28 dex for the shortest two FIR wavelengths, 0.43~dex for SPIRE \um{500}). The model is still clearly inferior to within-sample testing (Sect.~\ref{ssec-ml-general}), and there are quite some outliers besides the red points. 

As we will see in Sect.~\ref{ssec-model-interpretation}, the predictions for PACS tend to be mostly based on the MIR fluxes, and this estimate works well across data sets. The SPIRE predictions require a broader set of input broadbands, mostly relying on optical and NIR wavelengths. The relation to the input data is more complex, and so it is harder to extrapolate to a different data set. It is also clear that the predicted uncertainties are too large for the longest wavelengths: the uncertainty estimator is also inaccurate. 

We note that besides the plateauing of the predictions at a lower limit, there seems to be a general trend of overpredictions. Again, this bias increases towards longer wavelengths. When considering only galaxies with $y > 0$ (to avoid the plateau), we find a mean error of $\sim 0.24$~ dex (overpredictions) for the three SPIRE bands (and 0.03, 0.10, and 0.19~dex,  for PACS 70, 100, and \um{160} respectively). Since almost all of these galaxies are close to the H-ATLAS training set in feature space, we attribute this to the different data reductions of DustPedia and H-ATLAS. Differences include aperture matching (DustPedia uses a matched aperture for all bands, while H-ATLAS uses a different---typically smaller---aperture for the \textit{Herschel} bands), background subtraction, and foreground star removal. Part of the difference can also be explained by intrinsic differences in the data. We found that the SPIRE \um{500} to PACS \um{100} ratio in H-ATLAS is on average 0.1 dex higher than in DustPedia, suggesting that DustPedia contains galaxies with warmer dust. This could bias the predictions, but not enough to explain the 0.24~dex mean error.

Finally, we reversed the datasets; training on DustPedia and testing on H-ATLAS. This results in an overall RMSE of 0.36~dex (compared to 0.43~dex for H-ATLAS training, DustPedia testing). PACS \um{70} has a RMSE of 0.25 dex, and this gradually increases to a RMSE of 0.47~dex at SPIRE \um{500}. The lower RMSE is however due to the lower variance in H-ATLAS: at wavelengths longer than \um{160} the $R^2$ becomes negative. The bias is now reversed, since we underpredict on average. The plateau at longer wavelengths is also reversed, forming an upper limit on the predictions. 

\subsection{The importance of the MIR fluxes}
\label{ssec-nowise}

While the UV--NIR emission originates from stars, the WISE \um{12} and \um{22} bands are in the MIR and reveal the emission of small, hot dust grains. From the SEDs in Fig.~\ref{fig-example-seds}, we see that although this MIR is still distinct from the FIR, it does give a good hint of what to expect at longer wavelengths. We can hence expect that the MIR bands are important features. 

To assess this statement, we trained a model without using the WISE \um{12} and \um{22} bands. To properly evaluate the lack of these bands, we redid the SED modelling step (acquiring Bayesian estimates) for the input features. After training the models, we found a total RMSE of 0.27~dex (up from 0.19~dex). This shows that the MIR is an important part of our predictor, although we still significantly outperform the SED fitting approach (0.38~dex). 

The worst predictions are now for the shortest FIR wavelengths: PACS 70 and \um{100} have a RMSE of 0.34~dex and 0.29~dex, while the remaining bands have a RMSE around 0.23~dex. The full predicted vs true plot is shown in Appendix~\ref{app-figures}, Fig.~\ref{fig-truevspred-nnet-no12p}. The dust luminosity can be predicted with a RMSE of 0.26~dex (up from 0.16~dex), while the dust temperature has a RMSE of 3.49 K ($R^2 = 0.27$, so not much better than the baseline of 4.09 K). When leaving out the MIR, the additional complexity of neural networks is especially clear: a linear model has an overall RMSE of 0.33~dex, with all bands having an RMSE above $\sim 0.30$~dex. 

Next, we leave out all four WISE bands. Besides redoing the SED fitting, we also change the feature normalisation band to 2MASS $K_s$ (since \um{3.4} is no longer constrained by observations). The overall RMSE is now 0.29~dex, slightly worse than leaving out only \um{12} and \um{22}. The corresponding predicted vs true plot is shown in Fig.~\ref{fig-truevspred-nnet-nowise}. A linear model has an overall RMSE of 0.35~dex, so again more complexity helps.

\subsection{Understanding the data}
\label{ssec-data-interpretation}

To understand the model, it is first important to understand the data. We can investigate which galaxies are bright in the FIR (compared to the NIR), and if these can be identified from a few UV--MIR features. By looking at correlations between the input features and output target, this can be investigated in a model-agnostic way. For this, we use the Bayesian flux and property estimates from our SED fitting.

First, we investigate an optical colour, like $g - i$. It is known that such a colour is a good predictor of stellar mass-to-light \citep{bell-de-jong, zibetti2009}: bluer (typically spiral) galaxies tend to have younger stars, and hence a lower optical $M/L$ than the redder (often elliptical) galaxies. However, $g - i$ colour seems to have only a mild effect on the dust-to-stellar luminosity: the Spearman $\rho$ between the these two variables is only 0.094 (redder galaxies have a slightly higher $L_d / L_\star$). While dusty galaxies often have an intrinsically bluer spectrum (due to the young stars), the dust attenuation reddens the spectrum. The result is that an optical colour is not an effective far-infrared predictor. FUV - H (a typical specific SFR tracer), has an even lower correlation ($\rho = 0.016$).

We calculated the Spearman $\rho$ between all possible colours (of our 14 Bayesian UV--MIR fluxes) and $L_d / L_\star$. The colours that combine an optical flux with WISE \um{12} or \um{22} come out on top as the strongest correlators. For example, $i - \um{22}$ has the strongest correlation with $\rho = 0.92$. The dust emission from these MIR bands, although originating from a warmer dust component than the far-infrared, gives a clear hint of the total dust luminosity.

When omitting the \um{12} and \um{22} bands (also from the CIGALE fit, since they can have an influence on the Bayesian estimate of shorter bands), still most information seems to be towards the longer wavelengths. The strongest correlator with $L_d / L_\star$ is now $H - \um{3.4}$, with $\rho = 0.74$. When leaving out all the WISE bands, the strongest correlator is $H - K_s$, with $\rho = 0.53$. The UV seems to be particularly weak when used in a single colour: besides FUV - NUV ($\rho = 0.19$), all colours involving a GALEX band have $|\rho| < 0.14$.

\begin{figure}
	\centering
	\includegraphics[width=\hsize]{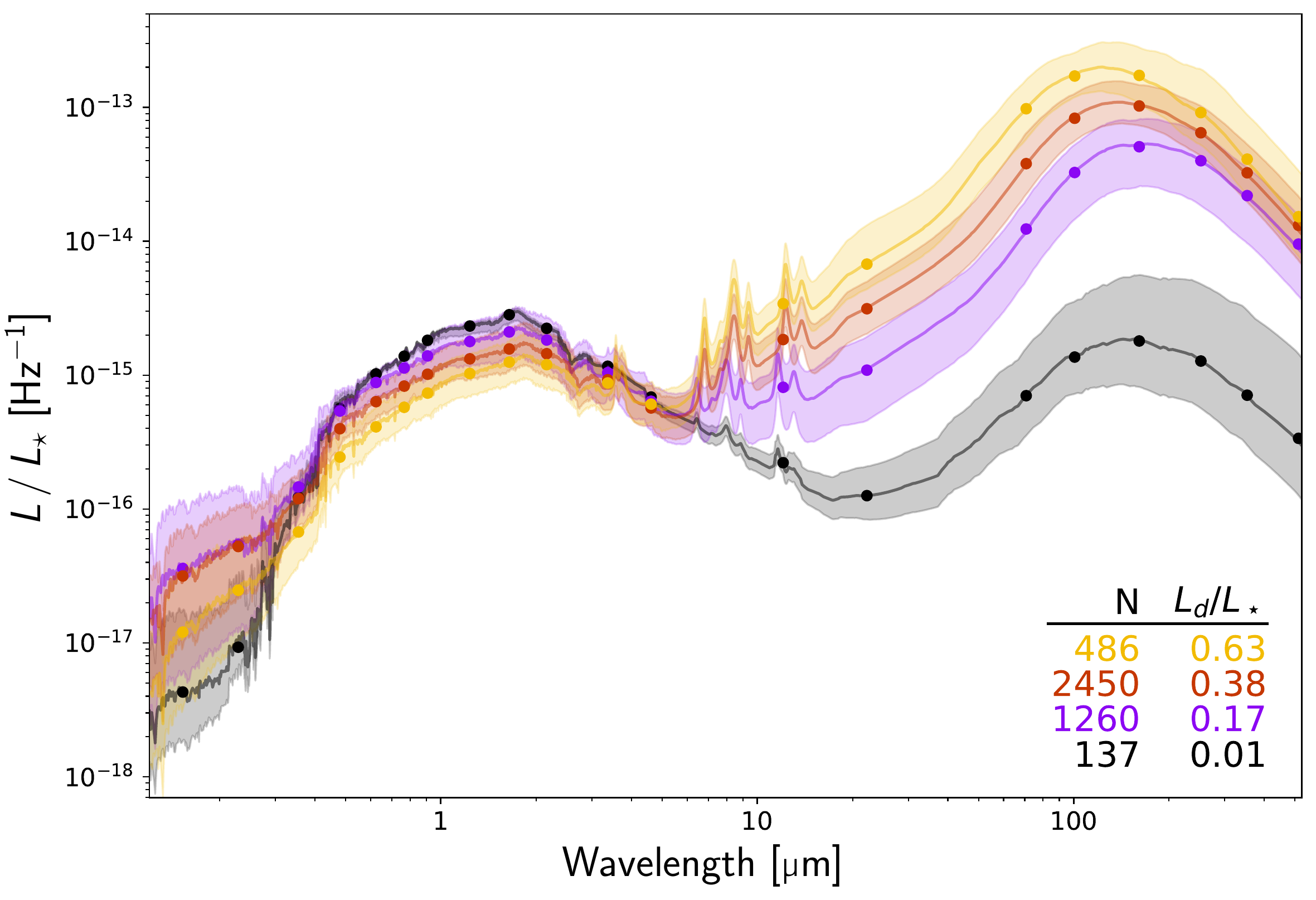}
	\caption{Sample divided in four bins of $L_d / L_\star$. The underlying SEDs are the Bayesian CIGALE estimates (UV--MIR CIGALE fit up to WISE \um{22}, UV--FIR CIGALE fit for longer wavelengths), and are normalised by the total stellar luminosity (from the short wavelength CIGALE fit). The markers show the median (per broadband) of each bin, and the shaded region is filled between the 16\textsuperscript{th} and 84\textsuperscript{th} percentile. These percentiles are taken for each Bayesian broadband flux. The CIGALE best models are used to interpolate between the broadband fluxes, in order to guide the eye. In the bottom right, the number of galaxies per bin and the median $L_d / L_\star$ per bin are shown.}
	\label{fig-seds-lfir-bin}
\end{figure}

Besides the dust luminosity, a second important property describing the FIR is the dust (effective) temperature. Warmer dust emits at shorter infrared wavelengths. Again, the longest wavelengths seem to be most important, with $\um{4.6} - \um{22}$ having the highest correlation with $T_d$ ($\rho = 0.45$). When there is a lot of \um{22} emission (w.r.t. \um{4.6}), the dust tends to be hotter: after all, \um{22} traces the warm dust. However, at the low \um{22} tail, there is again a slight upturn in cold dust temperature. Since Spearman's $\rho$ only quantifies monotonic relations, $\um{4.6} - \um{22}$ is a slightly better predictor of $T_d$ than might be expected from $\rho$ alone. Without \um{12} and \um{22} fluxes, the dust temperature seems to be very hard to constrain: $\rho(K_s - \um{3.4}) = 0.24$ is now the strongest correlation. The dust temperature correlates well with $L_d / L_\star$ ($\rho = 0.41$): warmer dust is more luminous (see Fig.~\ref{fig-seds-lfir-bin}, discussed next).

Besides the correlations with individual colours, we need to consider how the complete SED changes when varying $L_d / L_\star$. This is illustrated in Fig.~\ref{fig-seds-lfir-bin}. After ordering the sample by $L_d / L_\star$, we create four bins in dust-to-stellar luminosity (manually, balancing bin width and bin size). We show the median, 16\textsuperscript{th} and 84\textsuperscript{th} percentile of the SEDs of each bin, normalised by $L_\star$. From this, it quickly becomes clear why the NIR and MIR is so important to estimate the dust emission. The \um{12} and \um{22} give a good hint of the fluxes at longer wavelengths. There is a clear trend in the optical as well: dustier galaxies have less optical emission (normalised to the total stellar emission). This can be explained by the higher dust attenuation. Because the whole optical emission is offset, optical colours are not appropriate to estimate the FIR. However, the NIR (particularly \um{3.4} and \um{4.6}) provides an anchor point, since these bands correlate very well with the total stellar luminosity (which is dominated by the older stellar population). A colour combining optical and NIR is hence a better estimator of the FIR. 

The UV is spread over multiple orders of magnitude with no clear trend. It should be noted that the UV suffers from larger uncertainties: the uncertainties on the Bayesian flux are often of the same order of magnitude as the Bayesian fluxes themselves. Moreover, the UV is a more extreme regime, dominated by emission of luminous young stars, but heavily attenuated by both birth cloud and diffuse dust. 

\subsection{Model interpretation: derivatives}
\label{ssec-model-interpretation}

\begin{figure*}
	\centering
	\includegraphics[width=17cm]{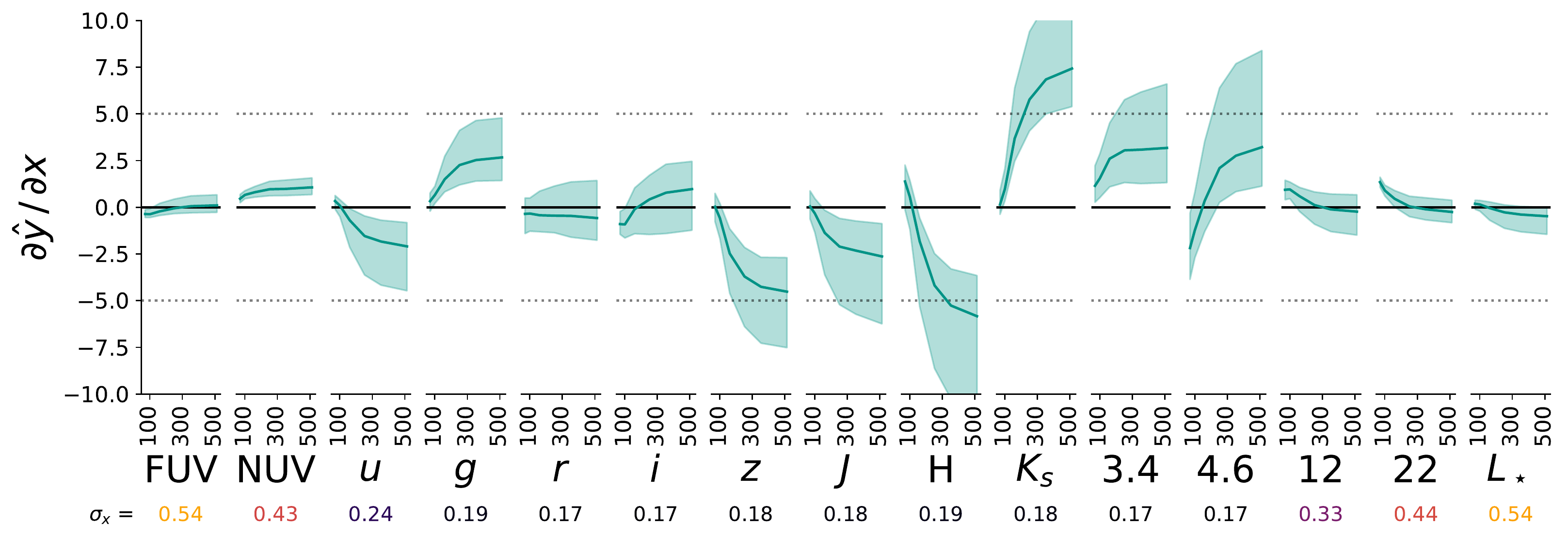}
	\caption{Partial derivatives of the model predictions with respect to the input features. Each panel shows one of the input features. Both features and target are normalised by $L_\star$ (except for $L_\star$ itself), and then logarithmised. Within each panel, we show the FIR components of the derivative (from \um{70} to \um{500} on a linear scale). The full line is the median over the whole data set (4~333 galaxies), while the shaded region shows 16\textsuperscript{th} and 84\textsuperscript{th} percentiles. The model consists of four neural networks, in a 4-fold train-test split. Under each input band, we show the standard deviation of the corresponding feature, coloured by its size.}
	\label{fig-feature-derivative}
\end{figure*}

Besides looking at relations from the data alone, we can use the model as a tool to investigate the relation between UV--MIR and FIR. Whereas the individual galaxies discretely sample the feature and target space, the model provides a smooth interpolation. It allows us to study variations in the output when travelling within feature space. One such variation is the partial derivative, of a particular target (FIR band) with respect to a particular input (UV--MIR band). This is shown in Fig.~\ref{fig-feature-derivative}. The partial derivative measures how much the prediction changes when moving a small amount along one feature. To ease interpretability, the model (4-fold train-test split) of Fig.~\ref{fig-feature-derivative} is normalised by $L_\star$ (instead of $F_{\um{3.4}}$), and again the log of this normalisation is used as an additional feature. This results in a slightly different model than our fiducial model (although the RMSE is essentially the same), which no longer places special focus on the \um{3.4} flux. The partial derivative is evaluated over the whole data set on each test fold, and the median, 16\textsuperscript{th}, and 84\textsuperscript{th} percentiles are shown.

First of all, it is quite remarkable how large the derivatives of some bands are. For $H$ and $K_s$, we get a more than $5\times$ response (at SPIRE wavelengths) to small inputs. However, we have to keep in mind that these adjacent broadbands do tend to vary together, and since their responses are opposite, they will tend to cancel out. It seems like the neural network translates these bands approximately into colours. Since the $H$ and $K_s$ responses are opposite, the $H - K_s$ colour response is boosted. The same is true for $u - g$. The Bayesian flux estimates are likelihood-weighted averages of the underlying SPS models, and hence these colours are fixed to the range of colours of the SPS models. Comparing to Fig.~\ref{fig-seds-lfir-bin}, we indeed see that the NIR acts as an anchor, with longer wavelengths correlating positively with FIR, and shorter wavelengths correlating negatively.

The PACS wavelengths vary less with input changes. \um{22} (and to a lesser extent \um{12}) has a consistent impact on PACS ($\times 1.3 \pm 0.3$ at PACS \um{70}), but on average little impact at the longest wavelengths. The relatively small impact of the \um{12} and \um{22} bands does not mean that they are not important (see Sect.~\ref{ssec-nowise}). Some bands---like \um{22}---are more likely to vary on their own (with fixed neighbouring bands) than for example $H$. This is why we give the standard deviation of each feature under the corresponding band name in Fig.~\ref{fig-feature-derivative}. The UV and MIR have the largest standard deviation: although their responses are lower, they are more likely to vary, hence increasing their importance. The Bayesian $\log{ \left( H / K_s \right)}$ varies only over 0.15 dex for the whole data set, and hence even a small variation can be significant, leading to the large responses. To determine the response of a standardised feature (i.e. rescaled to unit standard deviation), the given response $\partial y / \partial x$ should be multiplied with the feature's standard deviation $\sigma_x$. 

Overall, FUV and $L_\star$ seem to have little impact on the predictions. The largest effects seem to be from $i$ to $K_s$ (\um{0.76} to \um{2.16}). When we leave out GALEX, the overall RMSE increases only slightly (0.20~dex, up from fiducial 0.19~dex). So although NUV has a consistent response at longer wavelengths (making it important for the given neural network), a retrained model can perform well without this band. The $i$ to $K_s$ bands are most useful for the longer wavelengths, while WISE is most useful for PACS.

\subsection{Model interpretation: feature importance}
\label{ssec-feat-imp}

Besides the derivatives that we investigated in the previous section, there are other ways to measure the importance of a feature on the prediction. Random forests offer some commonly-used tools (such as selection frequency and Gini importance), but these can lead to unexpected results and are not applicable to our neural network models \citep{strobl2007}. A more intuitive method is permutation importance, which has the added benefit of being available for any predictor \citep{nicodemus2010}.

Permutation importance measures the influence of a feature on the predictor, by computing how the prediction performance changes after randomizing that feature. This randomisation is done by permuting all samples of a particular feature. Since the predictor is not retrained, this should decrease the performance, especially if the predictor heavily relies on that feature. If we measure the prediction performance by the RMSE, then the permutation importance is given by subtracting the standard (unpermuted) RMSE from the feature-permuted RMSE. Permutation importance measures a more global influence of a feature, whereas the derivatives of the previous section measure a local response. 

\begin{figure*}
	\centering
	\includegraphics[width=17cm]{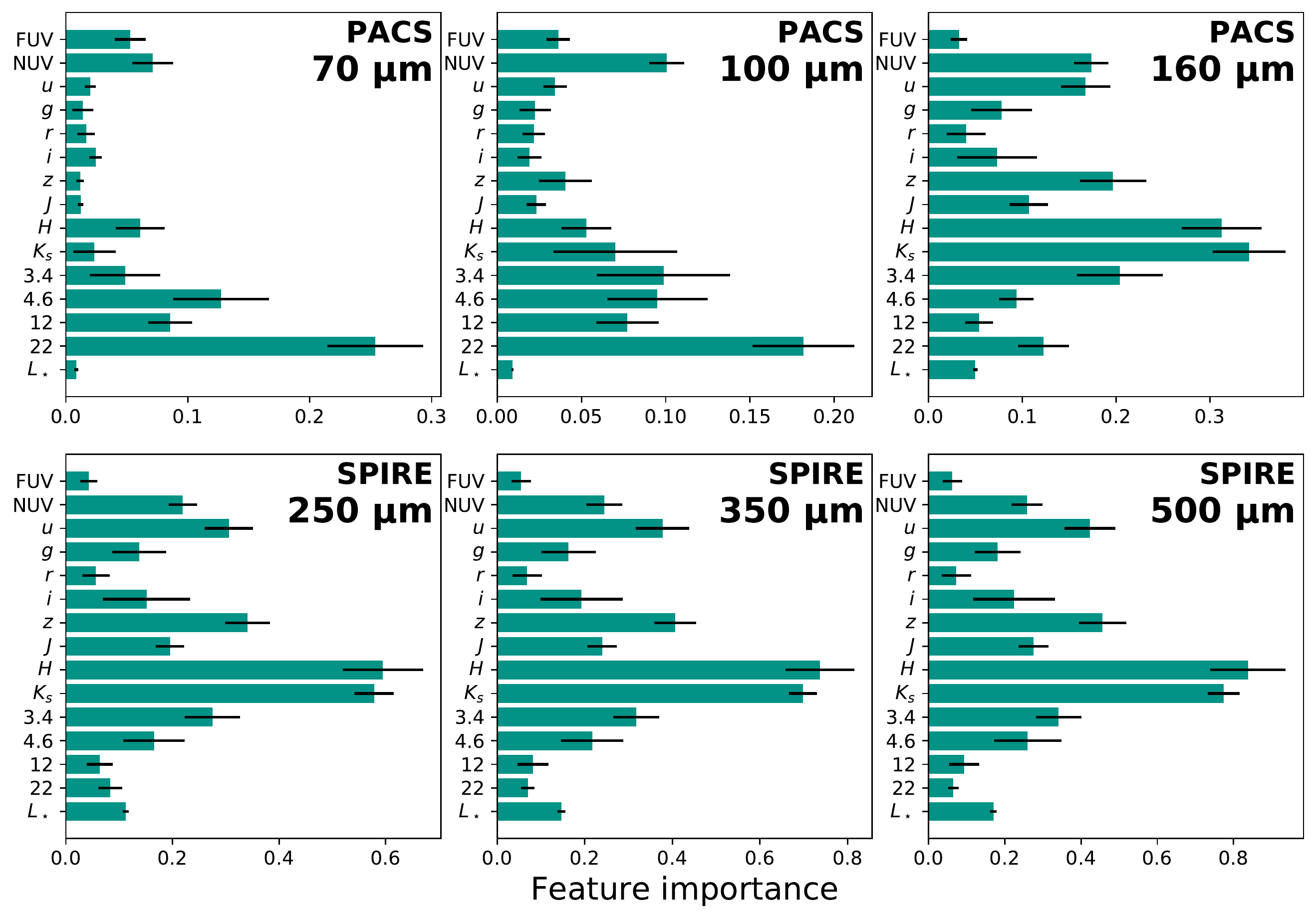}
	\caption{Permutation importance for the neural network regressor, normalised by $L_\star$. The average importance over the four train-test split models are shown, and the standard deviation is given as error bars. The importance of a feature is the RMSE of the model after randomizing that feature, minus the RMSE of the unpermuted predictions. }
	\label{fig-featimp-reg}
\end{figure*}

Fig.~\ref{fig-featimp-reg} shows the permutation importances for the neural network (normalised by $L_\star$). For PACS \um{70} and \um{100}, the \um{22} feature has the largest importance, while the optical has very little importance. When moving to longer FIR wavelengths, the importances shift more towards the NIR, peaking at the $H$-band. This matches well with what we found in the previous section: the \um{70} prediction relies mostly on the MIR, while \um{500} relies on the NIR.

Also at longer FIR wavelengths, NUV and $u$ are consistently more important than FUV and $g$. Overall, the stellar luminosity has not much influence on the predictions. However, it is worth noting that permuting a broadband feature leads to an unphysical SED (since it no longer aligns with neighbouring broadband fluxes), which is not the case for the stellar luminosity feature. This can reduce the importance of the $L_\star$ feature compared to the broadbands.

The same can be done for the uncertainty estimator, although here we can not use a RMSE metric. Instead, we use the negative log likelihood (on the test set), which was also used as the training loss. The results are shown in Appendix~\ref{app-figures}, Fig.~\ref{fig-featimp-unc}. Since the uncertainty estimator makes use of additional features, we only listed the top seven features, with most of the remaining features being consistent with zero importance. The \um{12} feature is the most important for every FIR band. FUV, \um{4.6}, and $L_\star$ are important for shorter, intermediate, and long wavelength FIR bands respectively. \um{22} and $F_{H, \mathrm{obs}} / F_{H, \mathrm{bay}}$ are also often included in the top features.

\section{Conclusions}
\label{sec-conclusions}

In this work we estimated the far-infrared SED, and its corresponding dust properties, via machine learning techniques. We used a low-redshift ($z < 0.1$) sample by combining DustPedia and H-ATLAS, both of which have UV to FIR data. Our input consisted of 14 broadband fluxes from  \um{0.15} to \um{22}. We processed both the input and output data using Bayesian SED fitting, normalised all bands to \um{3.4} and worked exclusively in log-space. We find following results:

\begin{itemize}
	\item Our method achieves a RMSE of 0.19~dex, significantly lower than a more traditional UV--MIR energy balance SED fitting approach (RMSE = 0.38 dex). Our predictions are also inherently unbiased for galaxies that are well represented by the training set. 
	\item The dust luminosity can be predicted especially well (RMSE = 0.16~dex), and can be used in SED fitting codes to constrain the dust attenuation without needing FIR. Predictions of dust mass have a RMSE of 0.30~dex, while (effective) dust temperature has more scatter (RMSE = 3~K).
	\item We can predict the uncertainty on our estimates, and this approach was validated using Fig.~\ref{fig-uncertainty-nnet}, which shows that the $\chi^2$ in each bin is consistent with 1.00.
	\item The worst predictions have either: 1) errors in their Bayesian \um{12} and/or \um{22} estimates (usually when these bands are not detected); 2) large uncertainties on the ground truth FIR; 3) an unusually large $F_{\um{22}} / F_{\um{70}}$ (large warm dust component). 
	\item Predicting cross-sample (e.g. training on H-ATLAS and testing on DustPedia) works reasonably, as long as we do not stray too far from the training set in feature space (RMSE = 0.35~dex when excluding the 100 galaxies that least resemble the training set). It is necessary to remove the total luminosity feature, to make the two data sets more similar.
	\item A predictor that did not have access to WISE \um{12} and \um{22} has RMSE = 0.26 dex (RMSE = 0.29 dex when leaving out all four WISE bands). Especially the predictions for the PACS bands deteriorate when WISE is not available.
	\item Bands between \um{2.1} and \um{4.6} can be used as an anchor point, fixing $L_\star$ (see Fig.~\ref{fig-seds-lfir-bin}). When shorter wavelengths (\um{0.7}-\um{3.4}) are more emissive than this anchor, we tend to have less FIR emission and colder dust. When \um{12} and \um{22} are more emissive, we have more FIR emission (especially for PACS). 
	\item The PACS \um{70} and \um{100} predictions strongly rely on the \um{12} and \um{22} inputs. To predict longer FIR wavelengths, broadband fluxes between $z$ and \um{4.6} proved to be the most reliable. Bands shorter than \um{0.7} tended to be less valuable. 
\end{itemize}

While we have used 14 UV--MIR broadband fluxes, our model can be retrained for a different set of broadbands. Since we use SED fitting to extract our fluxes, our approach can be used on a more inhomogeneous dataset, that observes different galaxies with a different set of filters. After all, the extracted broadbands need not match the broadbands that were used for the observations. In addition, spectra can be added to the input (possibly after a dimensionality reduction). Possible applications of our model are:

\begin{itemize}
	\item Studying the dust in large optical-MIR surveys
	\item Simulating detections of proposed FIR telescopes
	\item Constraining the total dust attenuation in SED fitting \citep[e.g.][]{gswlc-2, decleir2019}
	\item Deblending confused FIR sources \citep[e.g.][]{pearson2018}
\end{itemize}

\begin{acknowledgements}
W.D., S.V. and M.B. gratefully acknowledge support from the Flemish Fund for Scientific Research (FWO-Vlaanderen). W.D. is a pre-doctoral researcher of the FWO-Vlaanderen (application number 1122718N). Special thanks to the Flemish Supercomputer Centre (VSC) for providing computational resources and support. We gratefully acknowledge the support of NVIDIA Corporation with the donation of the Titan Xp GPU used for this research. Following open-source python projects were used extensively: scikit-learn \citep{scikit-learn}, skorch, pytorch, CIGALE, numpy, scipy, pandas, matplotlib. Many thanks to the maintainers and contributors of these projects.

DustPedia is a collaborative focused research project supported by the European Union under the Seventh Framework Programme (2007-2013) call (proposal no. 606824). The participating institutions are: Cardiff University, UK; National Observatory of Athens, Greece; Ghent University, Belgium; Université Paris Sud, France; National Institute for Astrophysics, Italy and CEA (Paris), France.
\end{acknowledgements}

\bibliographystyle{aa} 


\begin{appendix}

\section{Higher redshift}
\label{app-higherz}

In this appendix, we apply our model at higher redshifts. Two scenarios are considered. First, we train and test on a sample of H-ATLAS galaxies with $0.01 < z < 0.5$. Second, we train a model on $z < 0.1$ galaxies (as in the main text), but apply it to $0.1 < z < 0.5$. We also discuss the effect of K-correcting. For the underlying spectrum that a K-correction requires, we used the CIGALE best model fit.

\subsection{Higher redshift train and test}

\begin{figure*}
	\centering
	\includegraphics[width=17cm]{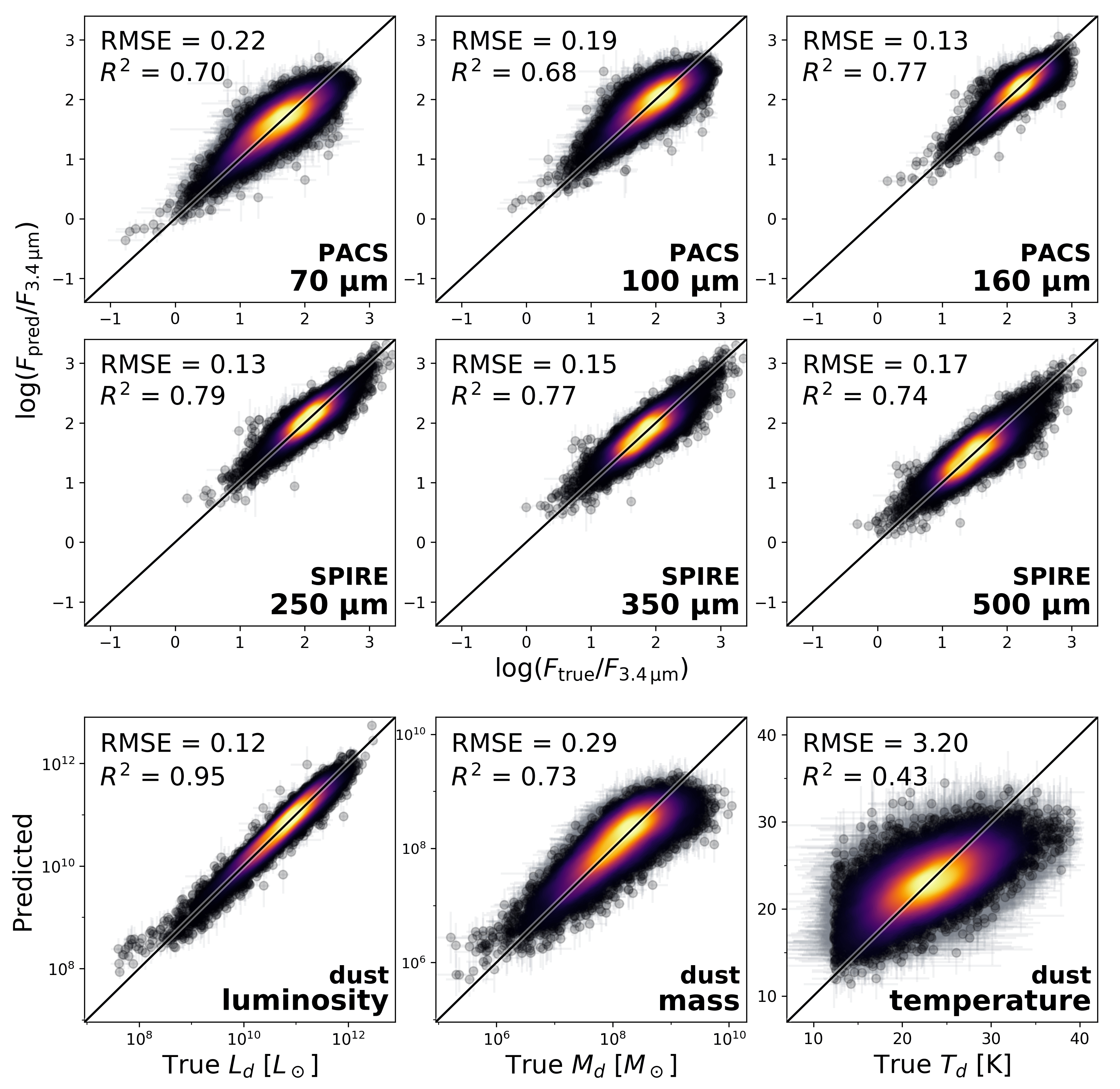}
	\caption{Similar to Fig.~\ref{fig-truevspred-nnet}, but for an extended redshift sample ($0.01 < z < 0.5$). The neural networks are both trained and tested on this redshift range (using a 4-fold train-test split). }
	\label{fig-truevspred-nnet-highz}
\end{figure*}

In Fig.~\ref{fig-truevspred-nnet-highz}, we show a predictor similar to our fiducial model (Sect.~\ref{sec-ml-results}), but using H-ATLAS data with an extended redshift range $0.01 < z < 0.5$. By again using a 4-fold train-test split, the total sample now contains 24~244 galaxies. When considering RMSE, the predictor performs similarly to our fiducial model: the overall RMSE is 0.170~dex, very close to the 0.171 dex of the $z < 0.1$ (H-ATLAS only) model. The $R^2$ is also similar to that of the H-ATLAS $z < 0.1$ model (about 0.08 lower for the shorter two wavelengths, but about 0.08 higher for the longest two wavelengths).

This larger sample also allows us to study the effect of sample size: larger training samples should lead to better performance. When training on 20\% of this sample (4~363 galaxies, similar to our fiducial model), the test performance only degrades by 0.006 dex. In other words, the sample size seems to be large enough to have reached convergence. With 10\% of this sample (2181 galaxies), the test performance degrades by 0.012 dex. With less galaxies, the performance quickly degrades. It seems that about 2~000 galaxies are needed for a satisfying performance using our shallow networks.

The predictor of Fig.~\ref{fig-truevspred-nnet-highz} did not use the redshift feature and was not K-corrected. We found that in this case, adding the redshift feature did not improve the results. However, when K-correcting, adding redshift as a feature notably improves the predictions (RMSE = 0.176 with redshift, RMSE = 0.201 without). The redshift is necessary for doing the K-correction, so it can as well be used as an additional feature. Still, K-correcting seems to slightly degrade the results. WISE \um{22} is an important feature to predict the far-infrared. For higher redshift galaxies, this band probes a shorter wavelength (\um{14.7} for $z = 0.5$). Recovering the rest-frame \um{22} band requires an extrapolation of the spectrum towards longer wavelengths -- which is exactly the goal of this work. The UV-MIR best model fit provides this extrapolation, but it is not very accurate (see Sect.~\ref{ssec-ml-general}), and hence results in an uncertain estimation of the rest-frame \um{22}. Adding the redshift feature allows the neural network to rely less on the K-corrected \um{22} feature for higher redshift galaxies. While the long wavelength part of the input is being extrapolated, the observed GALEX FUV is essentially discarded at $z = 0.5$ (since the rest-frame FUV is measured by the observed NUV).

K-correcting of course leads to a more apples-to-apples comparison. At redshift 0.43 the PACS \um{100} band actually measures a rest-frame PACS \um{70}. It is maybe surprising that an uncorrected predictor, without redshift feature does so well. The distance to the galaxy was however still incorporated in the WISE \um{3.4} luminosity: due to Malmquist bias, far away galaxies must be luminous. When leaving out this $L_{\um{3.4}}$ feature, we get a distance-agnostic predictor. This predictor has an overall RMSE of 0.202 dex. When leaving out the luminosity and redshift features, a K-corrected predictor performs slightly worse (RMSE = 0.217). Although we predict a fixed rest-frame wavelength, the K-correction leads to additional uncertainties that degrade the performance. For our fiducial model, leaving out the $L_{\um{3.4}}$ feature results in an overall RMSE of 0.221~dex (up from 0.194~dex). This is only a small increase in RMSE, and has the important benefit that we no longer need a spectroscopic redshift.

\subsection{Extrapolating to higher redshift}

\begin{figure*}
	\centering
	\includegraphics[width=17cm]{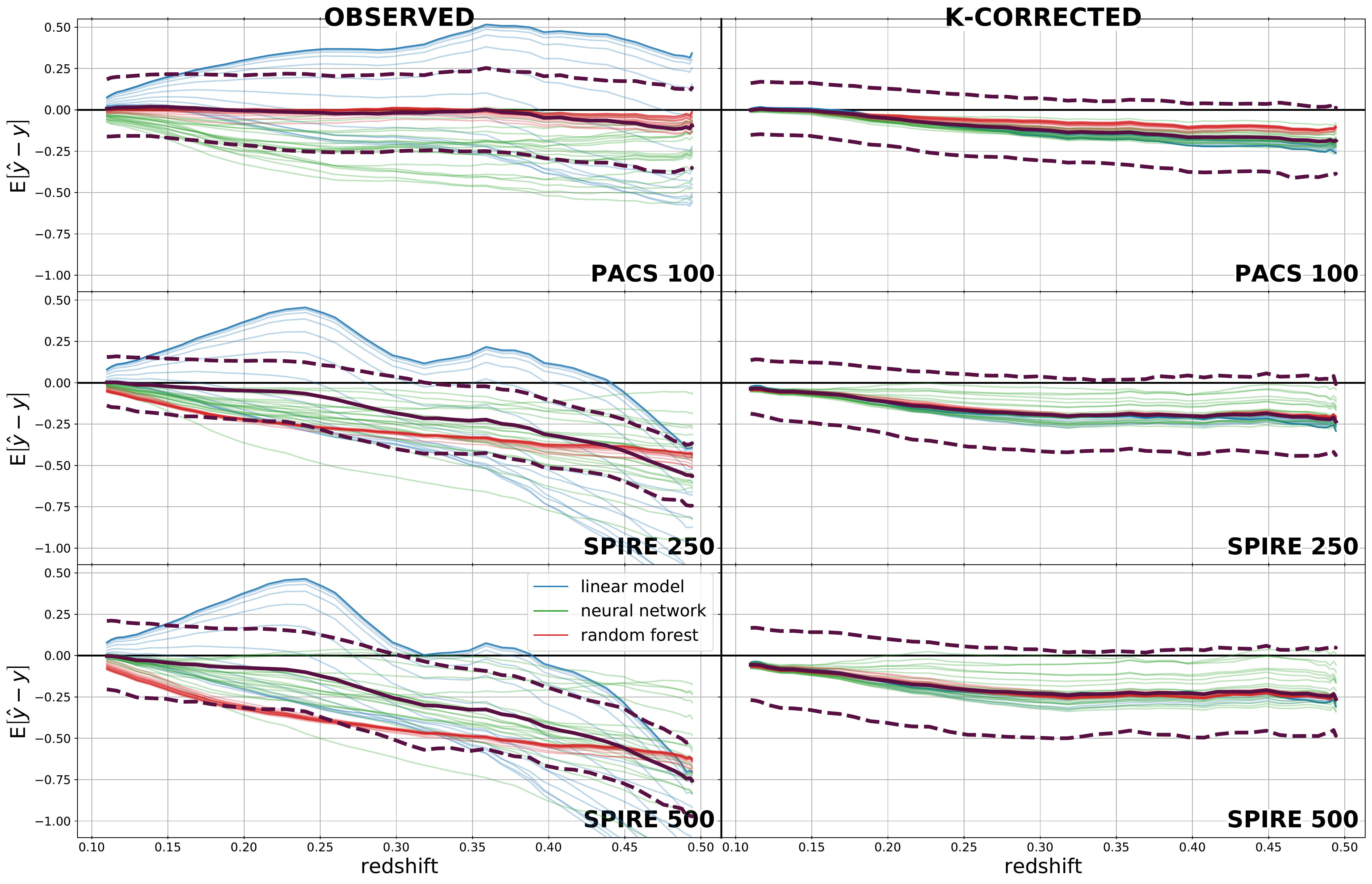}
	\caption{Mean error E$\left[\hat{y} - y\right]$ as function of redshift, for 60 models (per column). The three rows show PACS \um{100}, SPIRE \um{250} and SPIRE \um{500}. The left column uses uncorrected fluxes, while the right column is K-corrected. The models were trained on redshifts below 0.1 (DustPedia + H-ATLAS). An ensemble of the 60 models is shown with the dark plum lines, with the full line being the mean, and the dashed lines being the 16\textsuperscript{th} and 84\textsuperscript{th} percentiles.}
	\label{fig-highz-extrapolation}
\end{figure*}

If one wants to apply this method to predict FIR fluxes, it will almost certainly be on a different data set. DustPedia and H-ATLAS are useful because they have UV--FIR data, but the usefulness of our method is that it can be applied to a UV--MIR data set. Different samples can contain different biases. For example, a deeper sample would contain fainter galaxies that are not in our training set, and these should be used with extra care (as discussed in Sect.~\ref{ssec-nowise}, the luminosity feature should be removed). Here, we try to investigate if our method can be applied to higher redshifts. 

In Fig.~\ref{fig-highz-extrapolation}, we show a single line per estimator: a predictor's mean error as a function of redshift. Three types of estimators are used: 20 linear models (with different $L_2$ regularisation), 20 neural networks (with different architectures and random initialisations), and 20 random forests (with different decision tree depths). Each estimator was trained on our $z < 0.1$ sample (4~333 galaxies), without luminosity feature (no spectroscopic redshift needed for the uncorrected). A first noticeable result is that there is a lot of difference between the models when not K-correcting (left column). The ensemble of models remains mostly unbiased to $z = 0.25$, but this would probably not be the case when using a different mix of models. At higher redshifts, we tend to underestimate the luminosities, which can be attributed to the higher FIR luminosity of higher redshifts galaxies \citep{dunne2011-dust-evolution}.

The K-correction leads to more consistent results, as could be expected. Up to z = 0.17, the bias is small. For higher redshifts, we tend to underestimate all bands by about 0.25~dex. When correcting for this bias, the predictor works very well: the variance of the ensemble's predictions (indicated by the dashed lines) barely grows with redshift. Of course, it is impossible to know how large this bias is a priori, so again one should be careful when applying the model to a different set of galaxies.

\section{Additional figures}
\label{app-figures}

\begin{figure*}
	\centering
	\includegraphics[width=17cm]{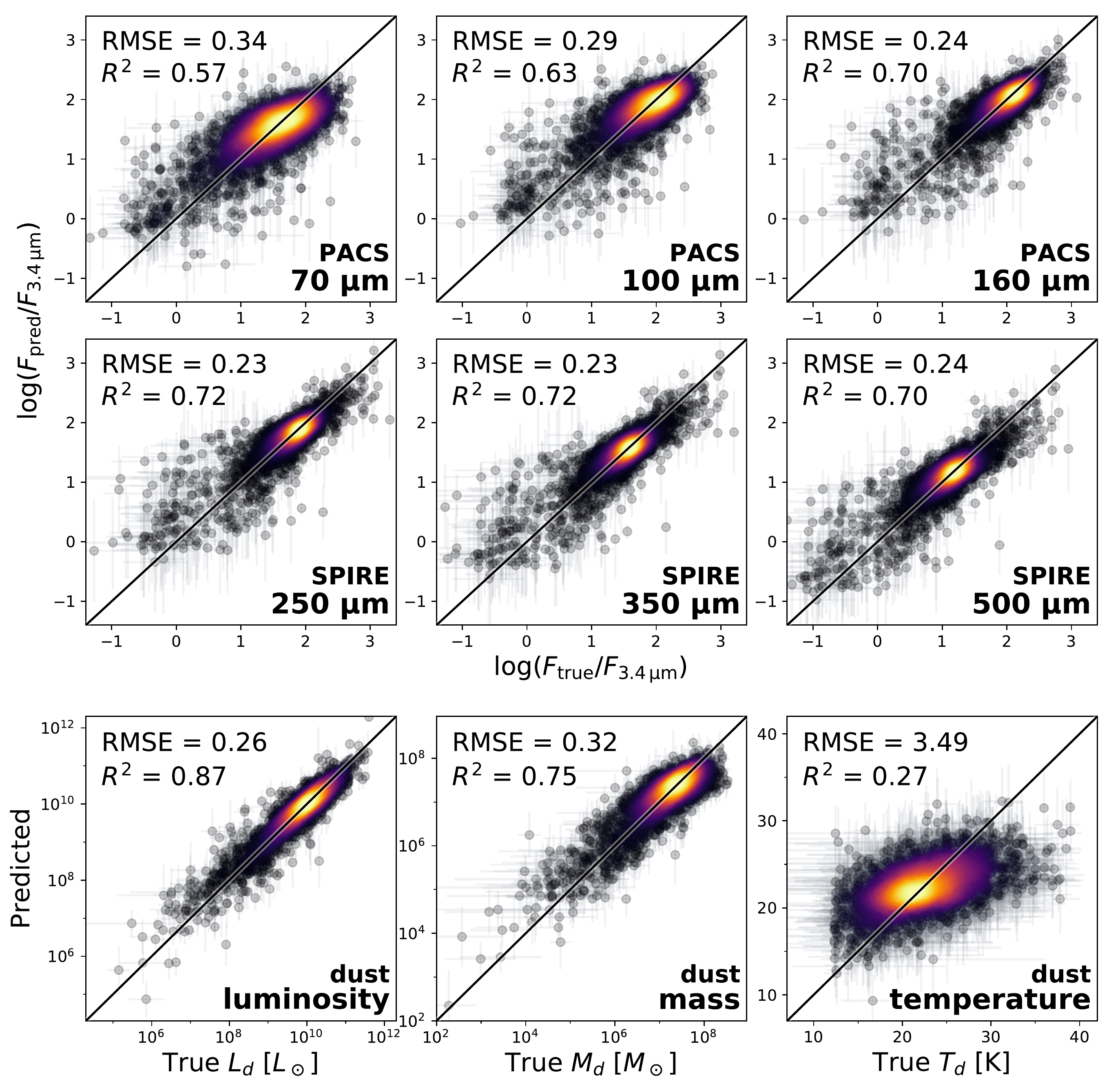}
	\caption{Similar to Fig.~\ref{fig-truevspred-nnet}, but for a model trained without WISE \um{12} and \um{22} bands. }
	\label{fig-truevspred-nnet-no12p}
\end{figure*}

\begin{figure*}
	\centering
	\includegraphics[width=17cm]{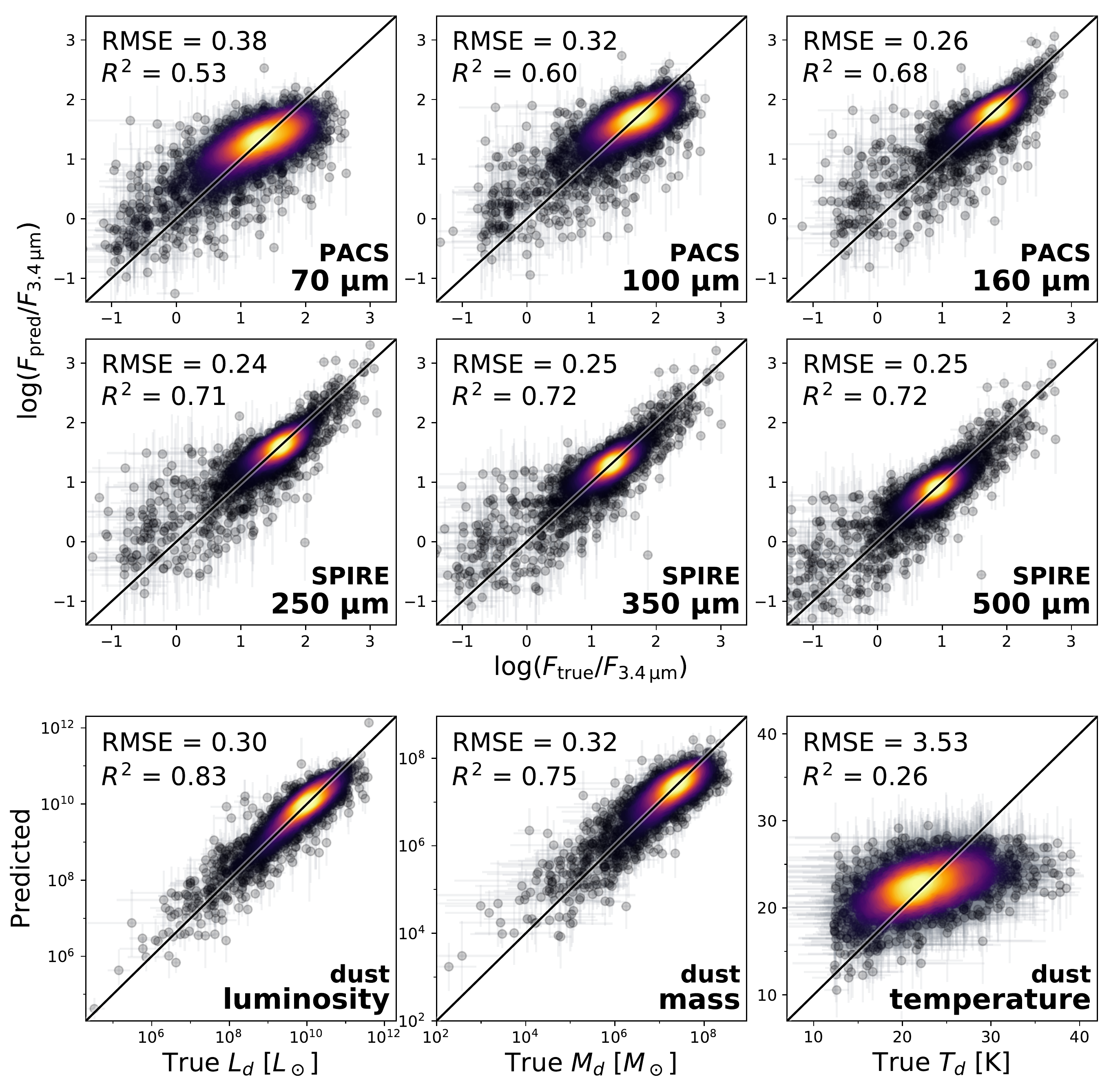}
	\caption{Similar to Fig.~\ref{fig-truevspred-nnet}, but for a model trained without all WISE bands (only GALEX, SDSS and 2MASS as input). }
	\label{fig-truevspred-nnet-nowise}
\end{figure*}

\begin{figure*}
	\centering
	\includegraphics[width=17cm]{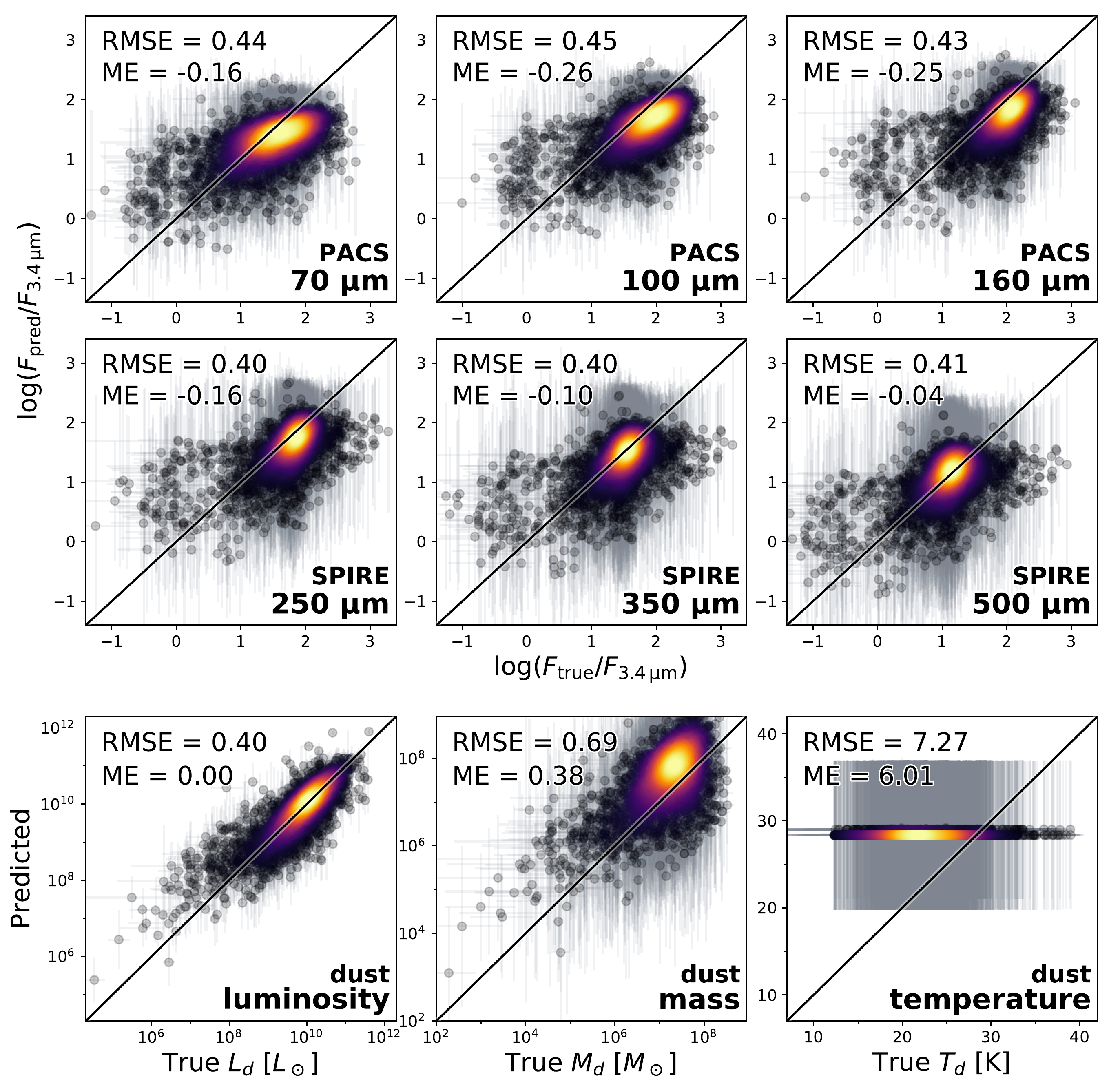}
	\caption{Similar to Fig.~\ref{fig-truevspred-cigale}, but leaving out the all WISE bands from the CIGALE fitting. }
	\label{fig-truevspred-cigale-nowise}
\end{figure*}

This appendix contains additional figures which supplement the main text. In Sect.~\ref{ssec-nowise}, we discuss models that do not make use of the MIR. Fig.~\ref{fig-truevspred-nnet-no12p} excludes WISE \um{12} and \um{22} from the input (also excluding them from the CIGALE fit). Fig.~\ref{fig-truevspred-nnet-nowise} excludes all WISE bands from the input (and CIGALE fit), and changes the normalisation band to 2MASS $K_s$. In addition, we show the UV-NIR CIGALE fitting predictions where all four WISE bands are omitted, as a function of the ground truth, in Fig.~\ref{fig-truevspred-cigale-nowise}. 

\begin{figure*}
	\centering
	\includegraphics[width=17cm]{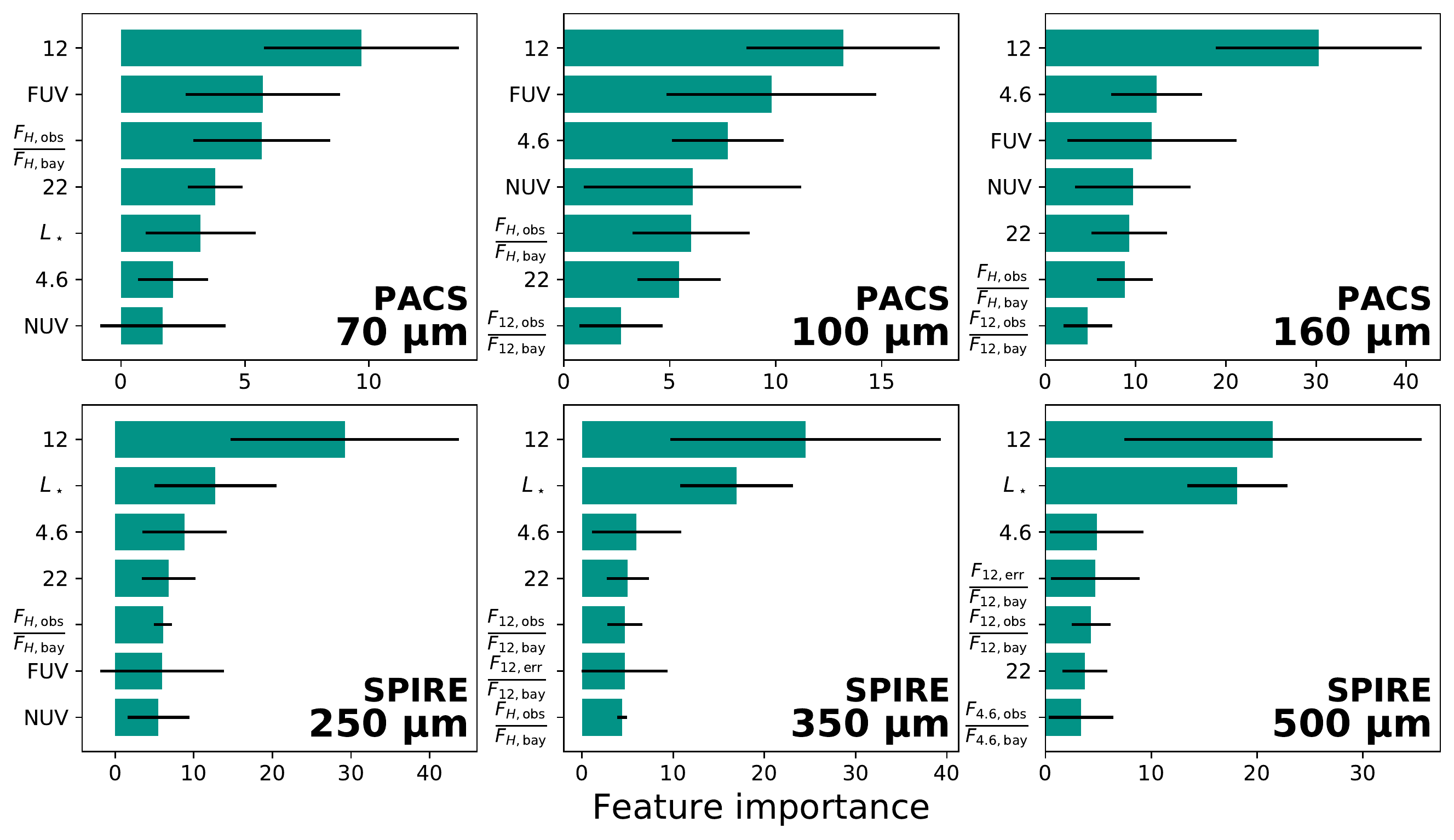}
	\caption{Similar to Fig.~\ref{fig-featimp-reg}, but for the uncertainty estimator (using the negative log likelihood metric). We only show the top seven features, since the remaining ones are almost all consistent with zero.}
	\label{fig-featimp-unc}
\end{figure*}

In Sect.~\ref{ssec-feat-imp}, we investigate the feature importance of our predictor. Fig.~\ref{fig-featimp-unc} shows the top seven permutation importances for the uncertainty estimator.

\section{Overview table}
\label{app-table}

\begin{sidewaystable*}[h]
	\caption{Overview of the RMSE of different machine learning models. All test sets are independent of the training sets. When the same sample is listed for the train and test set, 4 separate models are trained in a 4-fold train-test split (see Sect.~\ref{ssec-fluxpred}). }\label{tab-overview}
	\centering
	\begin{tabular}{@{\extracolsep{4pt}}ccccccccccc}
		\hline\hline
		\noalign{\smallskip} 
		&&\multicolumn{2}{c}{\it Sample} &\multicolumn{7}{c}{\it RMSE} \\
		\cline{3-4}  \cline{5-11} 
		\noalign{\smallskip}  Input & Model & Train & Test &\um{70} & \um{100} & \um{160} & \um{250} & \um{360} & \um{500} & total \\
		\hline
		\hline
		UV--MIR (14) & neural network & mixed & mixed & 0.22 & 0.19 & 0.17 & 0.18 & 0.20 & 0.21 & 0.20\\
		UV--MIR (14) & random forest & mixed & mixed & 0.22 & 0.19 & 0.18 & 0.20 & 0.22 & 0.24 & 0.21 \\
		UV--MIR (14) & linear regression & mixed & mixed & 0.23 & 0.21 & 0.21 & 0.23 & 0.25 & 0.27 & 0.23 \\
		UV--MIR + redshift (15) & neural network & mixed & mixed & 0.21 & 0.19 & 0.16 & 0.17 & 0.19 & 0.20 & 0.19 \\
		UV--MIR, no \um{3.4} (13) & neural network & H-ATLAS & DustPedia & 0.30 & 0.33 & 0.41 & 0.47 & 0.50 & 0.53 & 0.43 \\
		UV--MIR (14) & neural network & DustPedia & H-ATLAS & 0.26 & 0.25 & 0.30 & 0.38 & 0.43 & 0.47 & 0.36 \\
		UV--MIR (14) & neural network & DustPedia & DustPedia & 0.29 & 0.27 & 0.27 & 0.28 & 0.29 & 0.30 & 0.28 \\
		UV--MIR (14) & neural network  & H-ATLAS & H-ATLAS & 0.20 & 0.17 & 0.13 & 0.14 & 0.16 & 0.18 & 0.16 \\
		SDSS--MIR (12) & neural network & mixed & mixed & 0.23 & 0.20 & 0.17 & 0.19 & 0.21 & 0.22 & 0.20 \\
		2MASS--MIR (7) & neural network & mixed & mixed & 0.25 & 0.22 & 0.20 & 0.23 & 0.25 & 0.27 & 0.24 \\
		
		\hline
		\hline
	\end{tabular}
\end{sidewaystable*}
	
\end{appendix}

\end{document}